\documentclass[prl,
amsmath,amssymb,twocolumn
]{revtex4}
\usepackage{amssymb}
\usepackage{amsmath}
\usepackage{graphics}
\usepackage{graphicx}
\usepackage{subfigure}
\usepackage{dcolumn}
\usepackage{bm}
\usepackage{hyperref}
\usepackage{enumerate}
\usepackage{xcolor}
\usepackage{xr}
\newcommand{\vx}{\mathbf x}
\newcommand{\vX}{\mathbf X}

\begin{document}

\preprint{APS/123-QED}

\title{Line Stretching in Random Flows}

\author{D. R. Lester}
\affiliation{School of Engineering, RMIT University, 3000 Melbourne, Australia}
\email{daniel.lester@rmit.edu.au}
\author{M. Dentz}
\affiliation{Spanish National Research Council (IDAEA-CSIC), 08034 Barcelona, Spain}
\date{\today}

\begin{abstract}
How finite-sized material lines stretch in chaotic (mono-scale) and turbulent (multi-scale) flows remains a central but unresolved problem that governs mixing, transport and reaction. We show elongation is controlled by a finite-sampling process balancing ensemble and temporal averaging that is mediated by particle dispersion. These results expose the rich dynamics of line stretching and compel reassessment of experimental data and models of fluid-borne phenomena.
\end{abstract}


\maketitle

Fluid deformation governs a myriad fluid-borne processes including mixing and transport~\cite{Dimotakis:2005aa,Haller:2015aa,Villermaux:2019aa}, stress development in complex fluids~\cite{Rivlin:1971aa}, droplet breakup~\cite{Stone:1994aa}, particle alignment~\cite{Voth:2017aa}, 
chemical reactions~\cite{Libby:1976aa} and biological activity~\cite{Tel:2005aa}. Deformation can fundamentally augment these processes~\cite{Aref:2017aa,Sreenivasan:2019aa} and the rate of deformation is critical to their prediction in turbulent or chaotic flows. Indeed, the mean and variance of the rate of stretching of line elements act as inputs to many models~\cite{Villermaux:2003ab,Tel:2005aa,Haller:2008aa,Aref:2017aa} of these phenomena. Despite its fundamental role, line stretching is not fully understood; for over half a century this process has been quantified via phenomenological models~\cite{Cocke:1969aa,Girimaji:1990aa,Duplat:2010ab,Kalda:2000aa,Villermaux:2019aa} of classical fluid mechanics.

An \emph{ab intio} analysis of line stretching in random flows is performed in this study. The rich dynamics of line stretching in chaotic and turbulent flows is uncovered and previously incompatible results are reconciled. These results call for a reassessment of experimental data and models of fluid stretching and fluid-borne phenomena.

\begin{figure}
\centering
\begin{tabular}{c c}
\includegraphics[width=0.42\columnwidth]{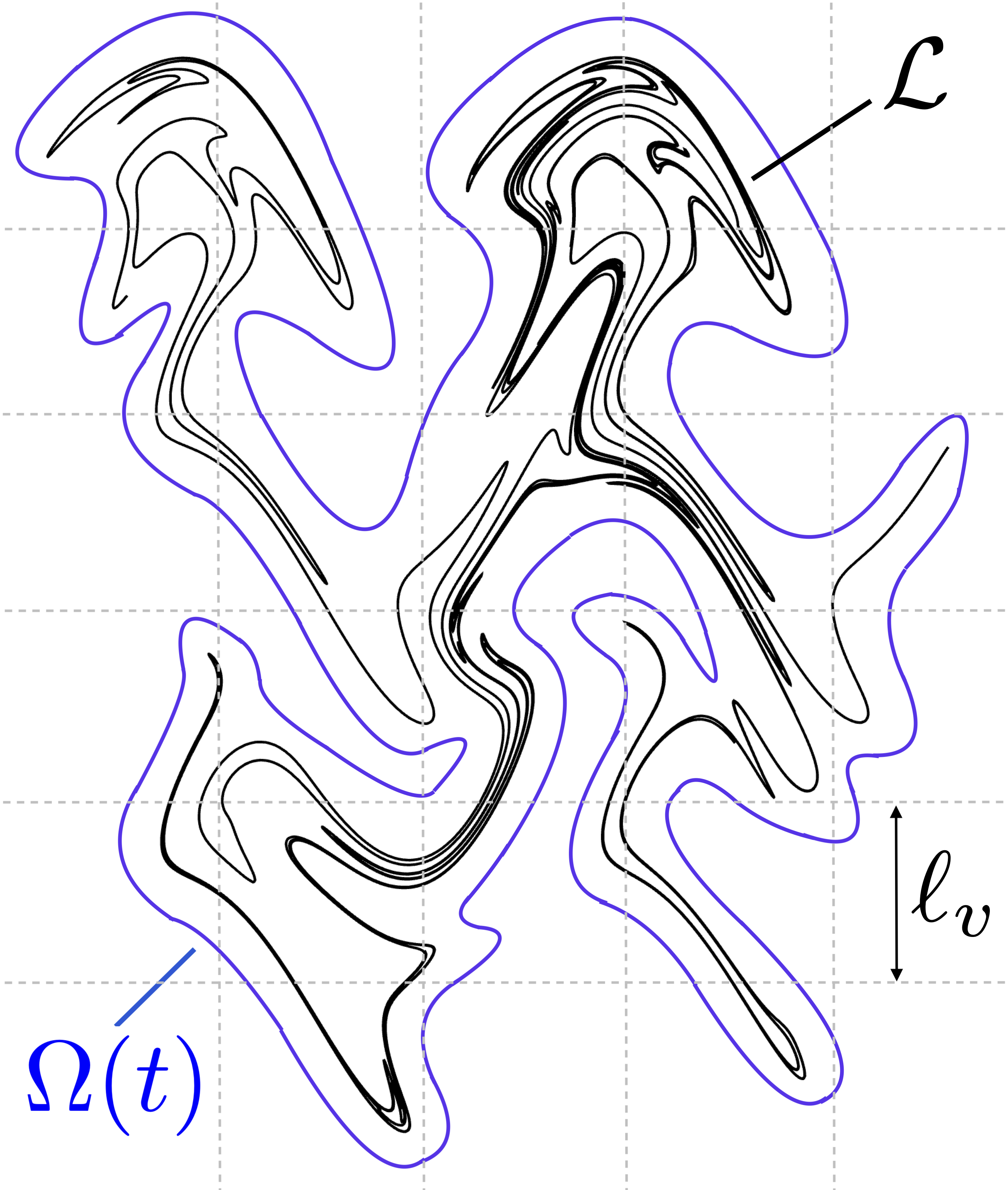}&
\includegraphics[width=0.57\columnwidth]{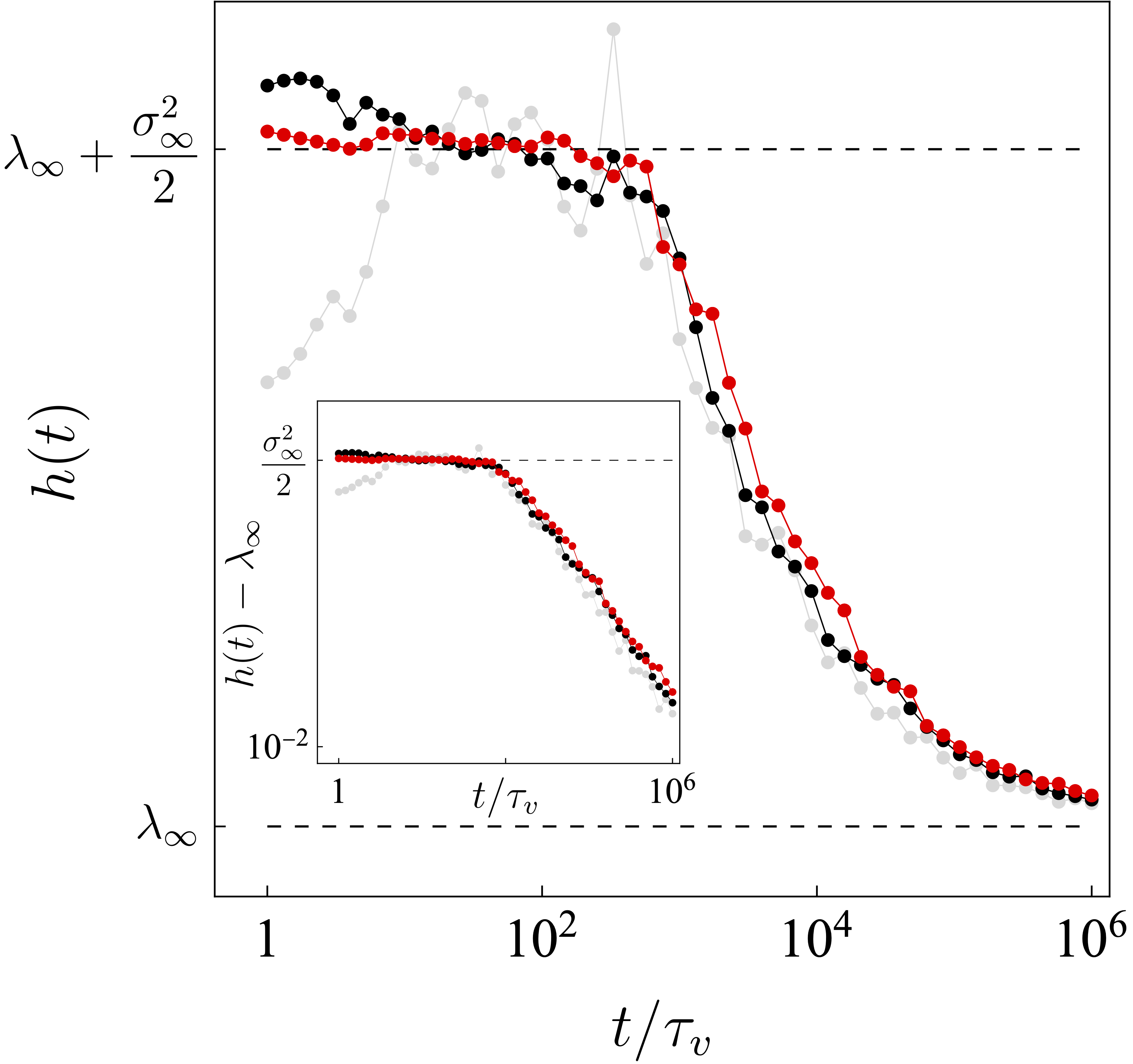}\\
(a) & (b)
\end{tabular}
\caption{(a) Stretching and folding of a material line $\mathcal{L}$ (black line) in a mono-scale flow over a coarse-grained finite support region $\Omega(t)$ (bounded by the blue line) that is discretised (dashed grey lines) with respect to $\ell_v$. 
(b) Evolution of $h(t)$ for homogeneous isotropic turbulence~\cite{Li:2008aa,Yeung:2002aa} ($Re_\lambda=433$) with Kolmogorov time $\tau_\eta = 0.0844$s~\cite{Li:2008aa} via statistical characterisation~\citep{Lester:2025aa} of stretching rates, with $\lambda_\infty=3.045$s$^{-1}$, $\sigma_\infty=\sigma_\epsilon\sqrt{\tau_v}=0.953$s$^{-1}$~\cite{suppmat}. Grey and black points respectively correspond to $h(t)$ for $N_0=10^3$, $10^4$  stretching variates, red points correspond to $N(t)$ based on turbulent dispersion~\cite{suppmat}. Inset details convergence of $h(t)$ to temporal average $\lambda_\infty$.}
\label{fig:plume}
\end{figure}

The two fundamental deformation measures are the Lyapunov exponent $\lambda_\infty$, which characterizes the asymptotic growth rate of infinitesimal line elements $\delta \ell$ as
\begin{equation}
 \lambda_\infty=\lim_{t\rightarrow\infty}\frac{1}{t}\ln\frac{\delta \ell(\mathbf{X},t)}{\delta \ell(\mathbf{X},0)}=\lim_{t\rightarrow\infty}\frac{1}{2t}\ln\nu(\mathbf{X},t),\label{eqn:lyapunov}
\end{equation}
(where $\nu$ is the largest eigenvalue of the Cauchy-Green tensor) and the topological entropy $h$ which measures the growth rate of the length $l(t)$ of finite material lines as
\begin{equation}
h=\lim_{t\rightarrow\infty}\left[h(t)\equiv\frac{1}{t}\ln\frac{\ell(t)}{\ell(0)}\right],\label{eqn:topo}
\end{equation}
where $h(t)$ is the finite-time topological entropy (FTTE). Some debate persists regarding these measures for purely hyperbolic flows. For \emph{mono-scale} flows with a single length scale $\ell_v$ (Fig~\ref{fig:plume}a), ergodic theorists~\cite{Matsuoka:2015aa,Newhouse:1993aa} show these are equivalent under certain conditions, i.e.
\begin{equation}
 h=\lambda_\infty,\label{eqn:entropy_mean}
\end{equation}
which represents a temporal average of stretching rates. Conversely, fluid physicists~\cite{Cocke:1969aa,Girimaji:1990aa,Kalda:2000aa,Goto:2002aa,Duplat:2010ab,Villermaux:2019aa} propose that stretching follows a random sequential process, leading to a Gaussian distribution of $\ln\delta \ell(t)$ with mean $\lambda_\infty\, t$ and variance $\sigma_\infty^2\,t$, and the ensemble average
\begin{equation}
    h=\lambda_\infty + \sigma_\infty^2/2.\label{eqn:entropy_var}
\end{equation}

The discrepancy between (\ref{eqn:entropy_mean}) and (\ref{eqn:entropy_var}) can be large, with $\sigma^2_\infty/2>\lambda_\infty$ in many turbulent~\cite{Meneveau:1991aa} and laminar~\cite{Lester:2025ac} flows. Although the stretching dynamics are more complex for \emph{multi-scale} flows such as turbulence, fluid stretching still follows a random sequential process~\cite{Girimaji:1990aa,Dresselhaus:1992aa,Tabor:1994aa,Goto:2002aa,Villermaux:2019aa}, and the competition between the ensemble and temporal averages persists. Indeed, Fig.~\ref{fig:plume}b shows initial convergence of $h(t)$ in homogeneous isotropic turbulence to (\ref{eqn:entropy_var}), followed by asymptotic convergence to (\ref{eqn:entropy_mean}).  This study explains these results via a theory of line stretching in mono- and multi-scale flows.

\emph{Fluid Deformation in Random Flows.} We consider stretching of a fluid element along a pathline $\mathbf{x}_0(\mathbf{X},t)$ with initial Lagrangian position $\mathbf{X}$ advected by the $d$-dimensional random uniformly hyperbolic velocity field $\mathbf{v}(\mathbf{x},t)$.
All such flows involve exponential growth of infinitesimal line elements $\delta\boldsymbol{\ell}(\mathbf{X},t)$. For simplicity we consider divergence-free flows, but note these results are easily extended to compressible flow. $\delta\boldsymbol{\ell}$ evolves according to the deformation gradient tensor $\mathbf{F}(\mathbf{X},t)\equiv\partial\mathbf{x}/\partial\mathbf{X}$ as
\begin{equation}
\delta\boldsymbol{\ell}(\mathbf{X},t) = \mathbf{F}(\mathbf{X},t)\,\delta\boldsymbol{\ell}(\mathbf{X},0),
\end{equation}
and $\mathbf{F}$ evolves via the Lagrangian velocity gradient tensor $\boldsymbol\epsilon(\mathbf{X},t)\equiv\nabla\mathbf{v}(\mathbf{x}_0(\mathbf{X},t),t)^\top$  as 
\begin{equation}
\dot{\mathbf{F}}(\mathbf{X},t)=\boldsymbol\epsilon(\mathbf{X},t)\,\mathbf{F}(\mathbf{X},t), \quad \mathbf{F}(\mathbf{X},0)=\mathbf{1},\label{eqn:deform}
\end{equation}
where $\det\mathbf{F}=1$ due to incompressibility. We denote the principal components of $\boldsymbol\epsilon$ in a $d$-dimensional flow as $\epsilon_i$ with $i=1:d$, where $\sum_{i=1}^d \epsilon_i=0$ and $\langle\epsilon_i\rangle\geqslant \langle\epsilon_{i+1}\rangle$. Due to rotation of material elements by vorticity and variation of $\boldsymbol\epsilon(\mathbf{X},t)$ with Lagrangian time $t$, line elements are not stretched according to $\epsilon_i$~\cite{Girimaji:1990aa,Dresselhaus:1992aa}. Rather, the quantities that govern line stretching are uncovered by rotating the Eulerian frame $\mathbf{x}$ into the \emph{Protean} coordinate frame $\mathbf{x}^\prime$~\cite{Lester:2018aa,Lester:2025aa} as
\begin{equation}
\mathbf{x}^\prime=\mathbf{x}_0(\mathbf{X},t)+\mathbf{Q}^\top(\mathbf{X},t)\,\mathbf{x},
\end{equation}
via the unitary rotation tensor $\mathbf{Q}(\mathbf{X},t)$, where the Protean frame is defined~\cite{Lester:2018aa,Lester:2025aa} by the rotation $\mathbf{Q}(\mathbf{X},t)$ that renders $\boldsymbol\epsilon^\prime(\mathbf{X},t)$ upper triangular. Hence $\mathbf{F}^\prime(\mathbf{X},t)=\mathbf{Q}^\top(\mathbf{X},t)\mathbf{F}(\mathbf{X},t)\mathbf{Q}(\mathbf{X},0)$ satisfies (\ref{eqn:deform}) with $\boldsymbol\epsilon\mapsto\boldsymbol\epsilon^\prime$, where
\begin{equation}
\boldsymbol\epsilon^\prime(\mathbf{X},t)=\mathbf{Q}\,
\boldsymbol\epsilon(\mathbf{X},t)\,\mathbf{Q}^\top
+\dot{\mathbf{Q}}^\top
\mathbf{Q}
\end{equation}

This transform has been applied to a range of flows, from isotropic homogeneous turbulence~\cite{Lester:2025aa} to laminar flows~\cite{Dentz:2016aa,Lester:2018aa} and recovers inherent kinematic constraints associated with non-chaotic flows~\cite{Lester:2022aa}. Due to rotation via $\mathbf{Q}$, the diagonal (principal) components $\epsilon_{ii}^\prime$ capture line stretching in both steady~\cite{Lester:2018aa} and unsteady~\cite{Lester:2025aa} flows, while the off-diagonal components $\epsilon_{ij}^\prime$, $i<j$ characterise shear. Hence the principal components $\epsilon_{ii}^\prime$ represent stretching rates, and the Lyapunov exponents of the flow are $\lambda_i=\langle\epsilon_{ii}^\prime\rangle$, with $\lambda_\infty\equiv\lambda_1>0$ and $\lambda_i\geqslant \lambda_{i+1}$. Due to the upper triangular form of $\boldsymbol\epsilon^\prime$, the diagonal elements of $\mathbf{F}^\prime$ then evolve via (\ref{eqn:deform}) as
\begin{equation}
F_{ii}^\prime(\mathbf{X},t)=\exp\left[\int_0^t \epsilon_{ii}^\prime(\mathbf{X},t^\prime)\, dt^\prime \right],\label{eqn:Fii}
\end{equation}
and so $F_{11}^\prime(\mathbf{X},t)$ governs growth of the largest eigenvector $\nu(\mathbf{X},t^\prime)$ of the Cauchy-Green tensor $\mathbf{C}\equiv\mathbf{F}^{\prime\top}\mathbf{F}^{\prime}=\mathbf{F}^{\top}\mathbf{F}$. For simplicity, the dominant stretching component $\epsilon_{11}^\prime$ is henceforth denoted $\epsilon$. Thus, the lengths of infinitesimal line elements $\delta l$ in (\ref{eqn:lyapunov}) grow according to the following multiplicative stretching process reported for a wide range of chaotic and turbulent flows~\cite{Villermaux:2019aa,Kree:2017aa,Souzy:2017aa,Villermaux:1994aa,Voth:2002aa,Heyman:2020aa}.
\begin{equation}
\delta \ell(\mathbf{X},t) \approx F^\prime_{11}(\mathbf{X},t) \delta l(\mathbf{X},0),\label{eqn:linestretch}
\end{equation}
For many random flows, the Lagrangian stretching rate $\epsilon'(t,\vX)$ resets randomly after a characteristic time scale $\tau_v$. For steady flows, this time scale is the same as that for the Lagrangian velocity~\cite{Le-Borgne:2008aa,Lester:2018aa}, but for unsteady flows such as turbulent flow~\cite{lumley:2007aa}, this is set by the Kolmogorov time scale, i.e. $\tau_v=\tau_\eta\equiv\sqrt{\nu/\varepsilon}$, with $\nu$ the kinematic viscosity and $\varepsilon$ the turbulent dissipation rate. Thus, the log-elongation $\xi(t) = \ln[\delta \ell(t,\vX)/\delta \ell(t,\vX)]$ has been modeled as a Brownian motion,
\begin{align}
    \frac{d \xi(t)}{dt} = \lambda_\infty + \sigma_\infty w(t),\label{eq:Wiener} 
\end{align}
where $w(t)$ is a Gaussian white noise and the variance $\sigma_\infty^2 = \sigma_\epsilon^2 \tau_v$. See~\cite{suppmat} regarding It\^{o} versus Stratonovic interpretation and approximation of Delta-correlated noise.

\emph{Line Stretching in Random Flows.} Equation \eqref{eqn:linestretch} quantifies stretching of line elements in random flows. For steady flows (which must be 3D to be hyperbolic), the Lyapunov spectrum $(\lambda_1,\dots,\lambda_d)$ is constrained to be $(\lambda_\infty,0,-\lambda_\infty)$, with null stretching in the velocity direction~\cite{Lester:2018aa}. For unsteady 2D flow, the Lyapunov spectrum is $(\lambda_\infty,-\lambda_\infty)$, where the stretching direction decouples from the velocity field, and for unsteady 3D flows, the Lyapunov spectrum is $(\lambda_\infty,\lambda_2,-\lambda_\infty-\lambda_2)$, with $\lambda_2>0$ consistently observed for turbulent flows~\cite{Pope:1994aa,Falkovich:2001ab}. As line elements do not align with the maximum stretching direction, $\lambda_\infty=\langle\epsilon\rangle$ is consistently smaller than $\langle\epsilon_1\rangle$~\cite{Girimaji:1990aa,Dresselhaus:1992aa,Tabor:1994aa}. 

For these hyperbolic flows, line elements $\delta l$ grow at an exponential rate~\cite{Girimaji:1990aa}, consistent with (\ref{eqn:linestretch}). The instantaneous stretching rate $\epsilon(\mathbf{X},t)$ is consistently observed~\cite{Yu:2023aa,Falkovich:2001ab,Lester:2025aa} to have finite mean $\lambda_\infty$ and variance $\sigma^2_\epsilon$. The Eulerian spatio-temporal correlation structure of $\mathbf{v}(\mathbf{x},t)$ also extends to the stretching rate $\epsilon(\mathbf{x},t)$~\cite{Lester:2018aa}. As shown in Fig.~\ref{fig:plume}a, the support volume $\Omega(t)$ for $\mathcal{L}$ evolves via a dispersion process, which plays a key role as it controls the sampling space of stretching rates $\epsilon$ that govern the evolution of $l(t)$. In the following we develop a model for evolution of $l(t)$ that honours both  processes.

Based on Eq. \eqref{eqn:linestretch}, the line elongation can  be written as the Lagrangian integral
\begin{align}
\label{eq:ell}
\ell(t) = \ell_0 \int\limits_{\Omega_0} d \vX \,\rho(\vX) \exp\left(\int\limits_0^t dt' \epsilon\left[\vx(t',\vX),t'\right]\right), 
\end{align}
where $\Omega_0$ is the volume comprising the initial line, $\rho(\vX) = \ell_0^{-1} \delta(X_1) \delta(X_2) \mathbb I(-\ell_0/2 \leq X_3 \leq \ell_0/2)$ denotes a uniform distribution of initial positions within a line of length $\ell_0$ along the $3$-direction of the coordinate system; $\mathbb I(\cdot)$ denotes the indicator function, which is one if the argument is true and zero otherwise. As the random stretching rate $\epsilon[\vx(t,\vX),t] \equiv \epsilon(t,\vX)$ fluctuates on the time scale $\tau_v$, we discretize expression \eqref{eq:ell} with respect to $\tau_v$ via
the path integral \cite{suppmat}
\begin{equation}
\label{eq:elln_markov}
    \begin{split}
        \ell_n = \ell_0 &\int d\vx_{n-1} \dots \int d\vx_0
g(\vx_{n-1},t_{n-1};\dots;\vx_0,t_0)\\
&\exp[\tau_v \epsilon(\vx_{n-1},t_{n-1})] \dots
\exp[\tau_v \epsilon(\vx_0,t_0)],
    \end{split}
\end{equation}
where $g(\vx_{n-1},t_{n-1};\dots;\vx_0,t_0)$ denotes the joint distribution of the positions of line elements that originate from the initial line due to dispersion of $\mathcal{L}$. Assuming that the $\vx(t|\vX)$ follow a Markovian dispersion process, the joint distribution decomposes into $g(\vx_{n-1},t_{n-1};\dots;\vx_0,t_0) = g(\vx_{n-1},t_{n-1}|\vx_{n-2},t_{n-2}) \dots g(\vx_1,t_1|\vx_0,t_0) g(\vx_0,t_0)$, where the $g(\vx_i,t_i|\vx_j,t_j)$ denote the conditional probability that a line segment is at position $\vx_i$ at time $t_i$ given that it was at position $\vx_j$ at time $t_j$. 

\emph{Zero Net Dispersion.} For illustrative purposes we first consider the simplest case of mono-scale flow with zero net advective dispersion that arises in \emph{isolated mixing regions} in chaotic flows~\cite{Bresler:1997aa}, where the line $\mathcal{L}$ is being exponentially stretched and folded within a bounded region of constant volume. For these flows, the line support $\Omega(t)$ has fixed volume $V_0$ that can be spatially discretised into \emph{stretching regions} of dimension $\ell_v \times \overset{d}{\dots}\times \ell_v$ (Fig.~\ref{fig:plume}a), such that $\mathcal{L}$ samples stretching rates that are independent between and uniform within each of the $N_0 \approx V_0/\ell_v^d$ stretching regions. Although the \emph{net} dispersion over $\Omega_0$ is zero, at each time step $\tau_v$, line segments within each stretching region are dispersed to several other stretching regions within $\Omega(t)$. 

For simplicity of exposition we assume for the time being that the distributions of line elements at a given time are independent of the distributions at all other times, that is, $g(\vx,t|\vx',t') = g(\vx,t)$, hence \eqref{eq:elln_markov} gives
\begin{align}
\label{eq:elln_markov0}
\ell_n = \ell_0 \prod\limits_{i = 0}^{n-1} \int d\mathbf{x}_{i} \exp[\tau_v
\epsilon(\mathbf{x}_{i},t_{i})] g(\mathbf{x}_{i},t_{i}),
\end{align}
Furthermore, we assume for the present that over each time step this dispersion within $\Omega(t)$ is \emph{complete}, such that at each time step the line segments in each stretching region are split into $N_0$ subsets and evenly distributed throughout $\Omega(t)$, hence the total number of line segments grows with time as $N_0^n$. Thus \eqref{eq:elln_markov0} may be discretized over these line segments to yield~\cite{suppmat}
\begin{align}
\label{eq:ell_ik}
\ell_n = \ell_0 \prod\limits_{i = 0}^{n-1} \frac{1}{N_0} \sum\limits_{k = 1}^{N_0} \exp(\tau_v \epsilon_{k,i}), 
\end{align}
where we used the fact that the support of $g(\vx,t)$ is always comprised of the same number $N_0$ of independent stretching regions. By definition, the $\epsilon_{k,i}$ are identical distributed random variables. Defining
$\epsilon_{k,i} \equiv \lambda_\infty + \epsilon'_{k,i}$, we can write
\begin{equation}
        \ell_n =  \frac{\ell_0 \exp(n\tau_v \lambda_\infty)}{\mathcal N_n} \sum\limits_{r=1}^{\mathcal N_n}
\exp\left(\zeta_{r,n}\right),
\label{eq:ell_zeta0}
\end{equation}
where $\mathcal{N}_n\equiv N_0^n$ and $\zeta_{r,n} \equiv \sum_{i = 1}^n \varepsilon'_{r,i} \tau_v$. The $\varepsilon'_{r,i}$ within a series $\{\varepsilon'_{r,i}\}_{i = 1}^n$ are mutually independent because for a given series the $\varepsilon'_{r,i} = \epsilon'_{k_i,i}$ with $k_i \in [1,N_0]$. As $\epsilon'_{k,n}$ have finite variance, then $\zeta_{r,n}$ are Gaussian distributed at times $t\gg \tau_v$. From \eqref{eq:Wiener}, we obtain that its mean is $\langle \zeta_{r,n} \rangle = 0$ and its variance is $\sigma_n^2 = n \sigma_\epsilon^2 \tau_v^2 \equiv n \tau_v \sigma_\infty^2$, and so \eqref{eq:ell_zeta0} can be written as 
\begin{align}
\label{eq:ell_zeta1}
\ell_n = \ell_0 \exp(n\tau_v \lambda_\infty) \int\limits_{-\infty}^\infty dx\, p_\zeta(x,n) \exp(x)
\end{align}
where we defined the one-point PDF of $\zeta_{r,n}$ as
\begin{align}
    p_\zeta(x,n) \equiv \frac{1}{\mathcal N_n} \sum\limits_{r=1}^{\mathcal N_n} \delta\left(z - \zeta_{r,n}\right).
\end{align}
Under the assumption of independent replicas $\zeta_{r,n}$, $p_\zeta$ converges to a Gaussian distribution as $\mathcal N_n\rightarrow\infty$ and so (\ref{eq:ell_zeta1}) recovers the ensemble average (\ref{eqn:entropy_var}) as $t\rightarrow\infty$. 
%
\begin{figure}
\centering
\begin{tabular}{c c}

\includegraphics[width=0.47\columnwidth]{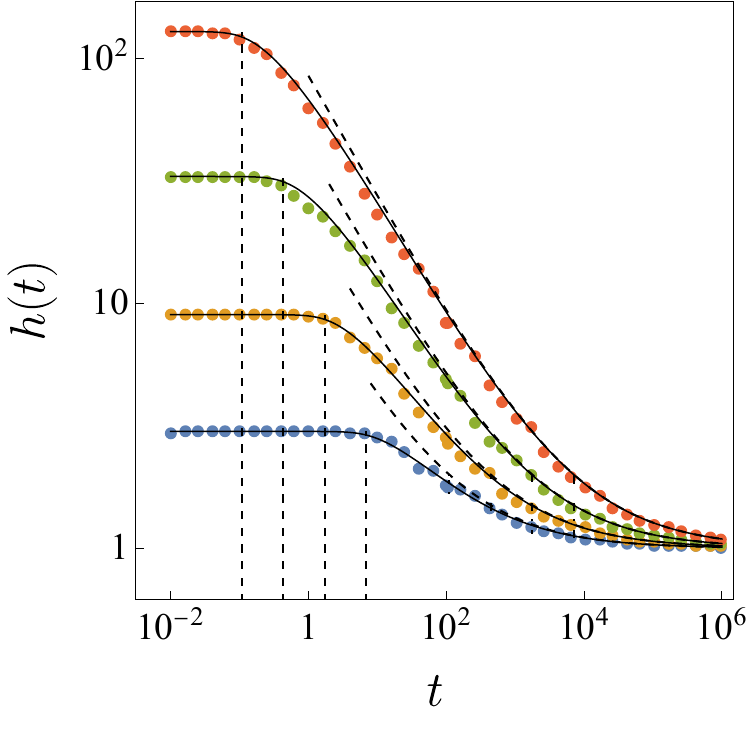}&
\includegraphics[width=0.48\columnwidth]{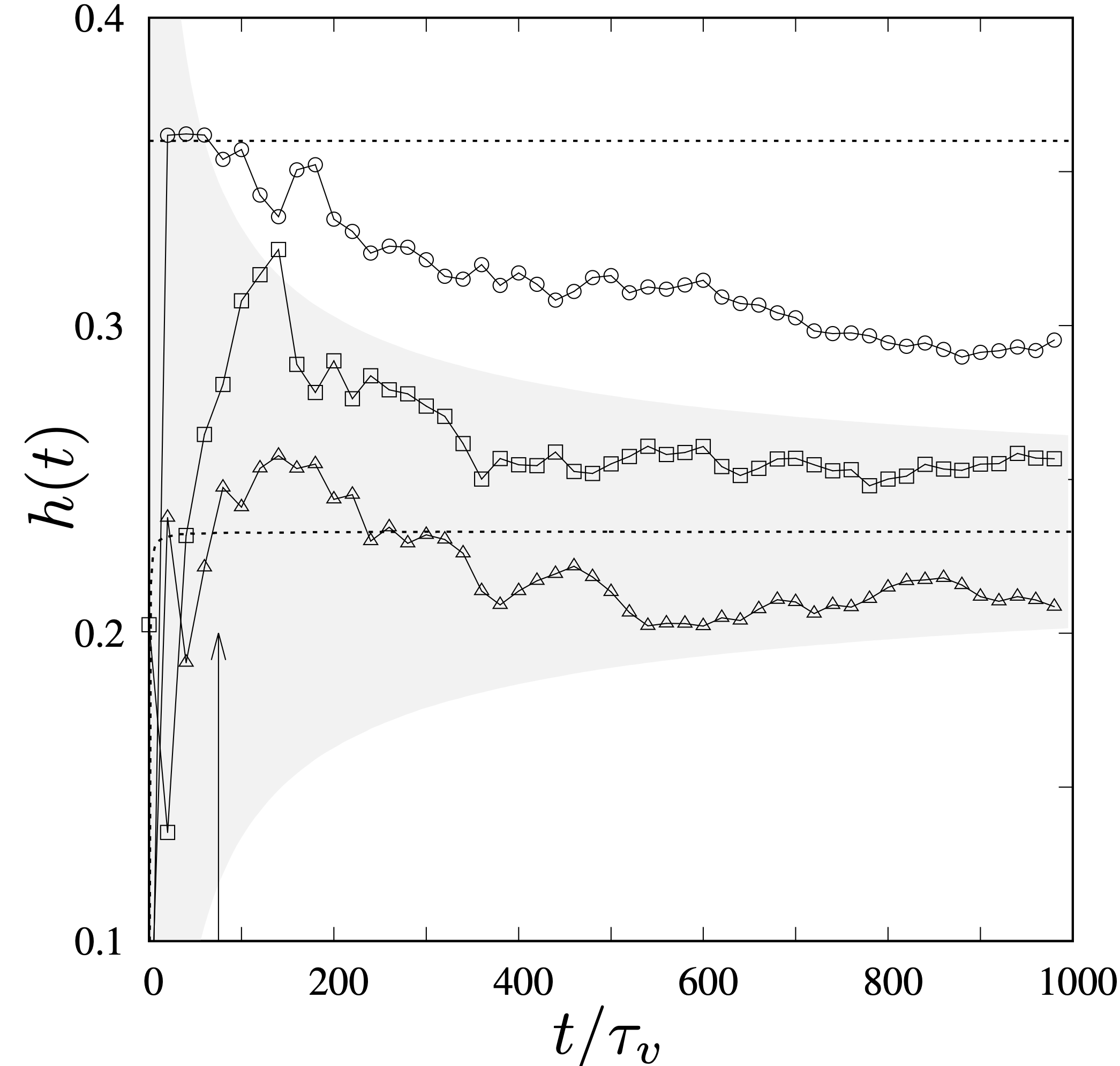}\\
(a) & (b)
\end{tabular}
\caption{(a) Convergence of $h(t)$ (dots) via statistical simulation of (\ref{eq:Wiener}) with $N(t)=10^6+10 t$ particles with $\lambda_\infty=1$ and $\sigma_\infty=2^m$, with $m=1:4$ (bottom to top)~\cite{suppmat}. Curved solid lines correspond to (\ref{eq:line_truncated}), curved dashed lines correspond to the asymptotic limit (\ref{eq:FTTE_4}), and vertical dashed lines correspond to time $t_1$ in (\ref{eqn:t1t2}). 
(b) Evolution of $h(t)$ for a material line $\mathcal{L}$ (left inset) in a 2D transient Kraichnan flow~\cite{Kraichnan:1970aa,suppmat} for (circles) ensemble average of $N_0 = 10^5$ infinitesimal line elements (square and triangles) individual line elements. The shaded area denotes the 95\% confidence interval for the finite-time Lyapunov exponent $\lambda_\infty\pm2\sigma_\infty/\sqrt{t}$. The arrow denotes $t_1 \approx 75 \tau_v$, the transition time given by (\ref{eqn:t1t2}).}
\label{fig:ctrw}
\end{figure}

However, although the number $\mathcal N_n$ of replicas of $\zeta_{r,n}$ increases exponentially with $n$, the number of \emph{independent} replicas does not due to the constant volume of the support $\Omega(t)$. Rather, the maximum number of independent replicas of $\zeta_{r,n}$ is equal to $n N_0/n = N_0$. Thus, the maximum $\zeta_{r,n}$ out of a sample of $\mathcal N_n$ dependent copies corresponds to the maximum out of a sample of $N_0$ independent copies. In \cite{suppmat} we demonstrate that for the turbulent flow considered in Fig.~\ref{fig:plume}b, $h(t)$ comprised of $N_0$ independent stretching variates $\epsilon$ at each time step is essentially identical to the \emph{replica} FTTE $h_r(t)$ comprised of a large number $\mathcal{N}_n\gg N_0$ of replicas $\zeta_{r,n}$ composed from this set of $\epsilon$. The distribution $p_\zeta(x,n)$ for a finite sample of $N_0$ replicas is then represented by the truncated Gaussian distribution~\cite{David:2011aa}
\begin{align}
\label{eq:pzeta}
p_\zeta(x,n) = \frac{1}{\text{erf}\left(\frac{A_{N_0} \sigma_n}{\sqrt{2}}\right)}\frac{\exp\left(- \frac{x^2}{2 \sigma_n^2} \right)}{\sqrt{2 \pi \sigma_n^2}}, 
\end{align}
for $|x| \leq A_{N_0} \sigma_n$ with
$A_{N_0} \sigma_n$ the expected value of the maximum in a sample of $N_0$ copies of $\zeta_{r,n}$. For large $N_0 \gg 1$ it can be approximated as $A_{N_0} \approx \sqrt{2 \ln N_0}$~\cite{David:2011aa}. Performing the continuum limit and setting $t = n \tau_v$, we obtain for the elongation 
%
\begin{equation}
\begin{split}
        \ell(t) = &\ell_0 
\frac{\exp\left(\lambda_\infty t+\frac{\sigma_\infty^2}{2} t\right)}{2\,\text{erf}\left(\frac{A_{N_0}}{\sqrt{2}}\right)}[\text{erf}\,\alpha_+(t)-\text{erf}\,\alpha_-(t)],
\label{eq:line_truncated}
\end{split}
\end{equation}
%
where $\alpha_\pm\equiv(\sigma_\infty \sqrt{t} \pm A_{N_0})/\sqrt{2}$. Fig.~\ref{fig:ctrw}a shows that this expression closely approximates numerical simulation of $h(t)$.

{\em Large time limit.} Defining
\begin{equation}
    t_1\equiv \frac{A_{N_0}^2}{\sigma_\infty^2}=\frac{2\ln N_0}{\sigma_\infty^2},\label{eqn:t1t2}
\end{equation}
then for $t \gg t_1$, the term in the square brackets can be approximated by $2 A_{N_0} \exp(-\sigma_\infty^2/2 t)\sqrt{2/\pi}$, hence in the long time limit,
\begin{align}
\ell(t) = \ell_0 \exp(\lambda_\infty t) A_{N_0} \sqrt{2/\pi}.  
\end{align}
and the topological entropy is given by the temporal average (\ref{eqn:entropy_mean}).

{\em Short time limit.} In the short time limit $t \ll t_1$, the expression in the square brackets in (\ref{eq:line_truncated}) can be approximated by $2 \text{erf}(A_{N_0})$, hence
\begin{align}
l(t) = \ell_0 \exp(\lambda_\infty t) \exp\left(\frac{\sigma_\infty^2}{2} t\right), 
\end{align}
which recovers the ensemble average (\ref{eqn:entropy_var}).

{\em Approaching the asymptotic limit.}
For large $N_0$ and long times, $t \gg t_1 \gg 1/\sigma_\infty^2$ we can approximate
\begin{align}
\ell(t) = \ell_0 \frac{\exp\left(\lambda_\infty t +\frac{\sigma_\infty^2}{2} t\right)}{\sqrt{2\pi}|\sigma_\infty \sqrt{t} - A_{N_0}|} \exp\left[-\frac{(\sigma_\infty \sqrt{t} - A_{N_0})^2}{2}\right].
\nonumber
\end{align}
From this expression, we obtain to leading order
\begin{align}
\label{eq:FTTE_3}
h(t) = 
\lambda_\infty + \sigma_\infty \sqrt{\frac{2 \ln N_0}{t}},
\end{align}
i.e., $h(t)$ converges to $h=\lambda_\infty$ as $1/\sqrt{t}$. As shown in Fig.~\ref{fig:ctrw}, $h(t)$ starts deviating from $\lambda_\infty+\sigma_\infty^2/2$ around $t_1$.

\emph{Highly Correlated Stretching.} Above we considered the extreme case where the distributions of line segments at subsequent time steps are independent. We consider now the opposite case that the line segments remain within the same stretching region at all times, that is, their positions are strongly correlated. In this case, we obtain \cite{suppmat}
\begin{align}
        \ell_n = \ell_0 \exp(n \tau_v \lambda_\infty) \frac{1}{N_0}\sum\limits_{k = 1}^{N_0} \exp\left(\zeta_{k,n}\right),
\end{align}
where $\zeta_{k,n} = \sum_{i = 1}^n \epsilon'_{k,n}$. Here, the $\zeta_{k,n}$ are independent for different $k$ and thus the number of independent replica is constant and exactly equal to $N_0$. Thus the FTTE evolves as discussed above. Cases involving partial dispersion over the support volume $\Omega_0$ and/or correlation between subsequent time steps lie between these two extremes and so behave in the same way.

\emph{Finite Net Dispersion.} For material lines that undergo finite net dispersion, the line support $\Omega(t)$ has evolving volume $V(t)$ that grows algebraically with time, as does the number $N(t)\approx V(t)/\ell_v^d$ of stretching variates. Again we assume that the distribution of stretchers at a given time step is independent from the distribution at the previous step, so the line elongation is given by \eqref{eq:elln_markov0}. For finite dispersion, the number of independent stretching regions denoted by $m_i$, that is, the number of independently distributed $\epsilon_{k,i}$, increases at each time step. Thus, we now discretize \eqref{eq:elln_markov0} as
\begin{align}
\label{eq:ell_ik2}
\ell_n = \ell_0 \exp(n \tau_v \lambda_\infty)\prod\limits_{i = 0}^{n-1} \frac{1}{m_i} \sum\limits_{k_i = 1}^{m_i} \exp(\tau_v \epsilon'_{k,i}), 
\end{align}
We can rewrite this expression in the form of Eq.~\eqref{eq:ell_zeta0}, where now $\mathcal N_n = \prod_{i=0}^{n-1} m_i$. 
It appears that the number of replica of $\zeta_{r,n} = \sum_{i = 0}^{n-1} \tau_v \varepsilon'_{r,i}$ increases exponentially fast with the number $n$ of time steps. However, the $\varepsilon'_{r,i}$ for a given $i$ are not independent, and therefore the $\zeta_{r,n}$ are correlated Gaussian random variables. The set of all independent $\epsilon'_{k,i}$ in Eq.~\eqref{eq:ell_ik2} after $n$ steps has the size $\sum_{i = 0}^{n-1} m_i$. Thus, the maximum number of independent replica $\zeta_{r,n}$ of length $n$ that can be formed is $N_n = \sum_{i = 0}^{n-1} m_i/n$. 
Following the same reasoning as above, the $\zeta_{r,n}$ form a finite sample of Gaussian random variables whose effective sample size is $N_n$. Thus, the expected value of the maximum of the sample of size $\mathcal N_n$ is equivalent to the one of a sample of $N_n$ members. Accordingly, the elongation follows \eqref{eq:ell_zeta0}. The only difference in \eqref{eq:pzeta} for $p_\zeta(x,n)$ is that $A_{N_0}$ is replaced by $A_{N_n}$. Following the same reasoning as above and substituting $N_0$ by $N_n$ in \eqref{eq:FTTE_3} gives for the FTTE
\begin{align}
\label{eq:FTTE_4}
h(t) = \lambda_\infty + \sigma_\infty \sqrt{\frac{2 \ln N(t)}{t}},
\end{align}
and the transition time $t_1$ in (\ref{eqn:t1t2}) can be updated according to $N_0\mapsto N(t)$. For algebraic growth of $V(t)$ and $N(t)$, which is typically observed for dispersion in random flows, topological entropy $h(t)$ still converges towards $\lambda_\infty$ as is shown in Fig~\ref{fig:ctrw}a, albeit slightly slower than the zero dispersion case ($N(t)=N_0$).

Fig.~\ref{fig:ctrw}b shows the stretching of a material line $\mathcal{L}$ in a mono-scale 2D Kraichan flow~\cite{suppmat}. As expected, the stretching rate of infinitesimal line elements all appear to converge toward $\lambda_\infty$, while $h(t)$ for an ensemble of $10^4$ particles appears to converge to the ensemble average (\ref{eqn:entropy_mean}) at times $t/\tau_v \leqslant t_1/\tau_v \approx 75$ before converging toward the temporal average (\ref{eqn:entropy_mean}) at longer times.

\emph{Multi-Scale Flows.} We now show that the results above also apply to multiscale flows such as homogeneous isotropic turbulence, which exhibit intermittency and scale-dependent statistics rather than the single scaling law predicted by classical Kolmogorov theory~\cite{Meneveau:1991aa,Kolmogorov:1991aa}. Such flows possess a broad spectrum of velocity scales and strongly non-Gaussian finite-time Lyapunov exponent (FTLE) statistics arising from intermittent vortex stretching~\cite{Meneveau:1991aa}. Nevertheless, stretching of a continuous material line is governed by the smallest flow scales and controlled by the Kolmogorov length $\eta\sim(\nu^3/\varepsilon)^{1/4}$ and time $\tau_\eta\sim\sqrt{\nu/\varepsilon}$~\cite{Kolmogorov:1991aa,lumley:2007aa}. This range of velocity length scales implies that $\Omega(t)$ partitions into stretching regions of varying volumes, which is captured by the above formulation, as discussed in \cite{suppmat}. While intermittency produces non-Gaussian FTLE statistics at intermediate times, the distribution converges to Gaussian for $t\gg\tau_\eta$~\cite{Ho:2020aa,Yu:2023aa}. Intermittency also drives anomalous dispersion $\langle\Delta r^2(t)\rangle\sim t^p$ with $1\le p\le3$, implying $V(t)\sim t^{dp/2}$. Together with ubiquitous folding in random flows, this yields algebraic growth of $N(t)$ such that $h(t)\to\lambda_\infty$, as shown in Fig.~\ref{fig:plume}b. 

\emph{Closing Remarks.} 
These results uncover the mechanisms governing the stretching of material lines $\mathcal{L}$ in mono- and multi-scale random flows, and more generally the deformation of finite-sized material elements. Dispersion controls the number $N(t)$ of independent stretching variates sampled by $\mathcal{L}$, yielding a competition between convergence of the finite-time topological entropy $h(t)$ to its ensemble (\ref{eqn:entropy_var}) and temporal (\ref{eqn:entropy_mean}) averages.

The rate of convergence depends on time, flow spatio-temporal structure, and dispersion, through their influence on the growth of $N(t)$, ultimately leading to convergence of $h(t)\rightarrow h=\lambda_\infty$. We thereby explain the commonly observed short-time convergence to the ensemble average ($t<t_1$) and reveal a transition to temporal averaging at long times ($t>t_1$). These findings enable inference of stretching dynamics directly from $l(t)$ data and call for a reassessment of experimental measurements and models of fluid deformation, with implications for mixing, colloidal deposition, and reactive transport in random flows.

\bibliography{reflist}

\begin{thebibliography}{2}%
\makeatletter
\providecommand \@ifxundefined [1]{%
 \@ifx{#1\undefined}
}%
\providecommand \@ifnum [1]{%
 \ifnum #1\expandafter \@firstoftwo
 \else \expandafter \@secondoftwo
 \fi
}%
\providecommand \@ifx [1]{%
 \ifx #1\expandafter \@firstoftwo
 \else \expandafter \@secondoftwo
 \fi
}%
\providecommand \natexlab [1]{#1}%
\providecommand \enquote  [1]{``#1''}%
\providecommand \bibnamefont  [1]{#1}%
\providecommand \bibfnamefont [1]{#1}%
\providecommand \citenamefont [1]{#1}%
\providecommand \href@noop [0]{\@secondoftwo}%
\providecommand \href [0]{\begingroup \@sanitize@url \@href}%
\providecommand \@href[1]{\@@startlink{#1}\@@href}%
\providecommand \@@href[1]{\endgroup#1\@@endlink}%
\providecommand \@sanitize@url [0]{\catcode `\\12\catcode `\$12\catcode
  `\&12\catcode `\#12\catcode `\^12\catcode `\_12\catcode `\%12\relax}%
\providecommand \@@startlink[1]{}%
\providecommand \@@endlink[0]{}%
\providecommand \url  [0]{\begingroup\@sanitize@url \@url }%
\providecommand \@url [1]{\endgroup\@href {#1}{\urlprefix }}%
\providecommand \urlprefix  [0]{URL }%
\providecommand \Eprint [0]{\href }%
\providecommand \doibase [0]{http://dx.doi.org/}%
\providecommand \selectlanguage [0]{\@gobble}%
\providecommand \bibinfo  [0]{\@secondoftwo}%
\providecommand \bibfield  [0]{\@secondoftwo}%
\providecommand \translation [1]{[#1]}%
\providecommand \BibitemOpen [0]{}%
\providecommand \bibitemStop [0]{}%
\providecommand \bibitemNoStop [0]{.\EOS\space}%
\providecommand \EOS [0]{\spacefactor3000\relax}%
\providecommand \BibitemShut  [1]{\csname bibitem#1\endcsname}%
\let\auto@bib@innerbib\@empty
\bibitem [{\citenamefont {Lester}\ and\ \citenamefont
  {Dentz}(2025)}]{Lester:2025aa}%
  \BibitemOpen
  \bibfield  {author} {\bibinfo {author} {\bibfnamefont {D.~R.}\ \bibnamefont
  {Lester}}\ and\ \bibinfo {author} {\bibfnamefont {M.}~\bibnamefont {Dentz}},\
  }\href@noop {} {\bibfield  {journal} {\bibinfo  {journal} {arXiv}\ }\textbf
  {\bibinfo {volume} {submit/6850609}} (\bibinfo {year} {2025})}\BibitemShut
  {NoStop}%
\bibitem [{\citenamefont {Li}\ \emph {et~al.}(2008)\citenamefont {Li},
  \citenamefont {Perlman}, \citenamefont {Wan}, \citenamefont {Yang},
  \citenamefont {Meneveau}, \citenamefont {Burns}, \citenamefont {Chen},
  \citenamefont {Szalay},\ and\ \citenamefont {Eyink}}]{Li:2008aa}%
  \BibitemOpen
  \bibfield  {author} {\bibinfo {author} {\bibfnamefont {Y.}~\bibnamefont
  {Li}}, \bibinfo {author} {\bibfnamefont {E.}~\bibnamefont {Perlman}},
  \bibinfo {author} {\bibfnamefont {M.}~\bibnamefont {Wan}}, \bibinfo {author}
  {\bibfnamefont {Y.}~\bibnamefont {Yang}}, \bibinfo {author} {\bibfnamefont
  {C.}~\bibnamefont {Meneveau}}, \bibinfo {author} {\bibfnamefont
  {R.}~\bibnamefont {Burns}}, \bibinfo {author} {\bibfnamefont
  {S.}~\bibnamefont {Chen}}, \bibinfo {author} {\bibfnamefont {A.}~\bibnamefont
  {Szalay}}, \ and\ \bibinfo {author} {\bibfnamefont {G.}~\bibnamefont
  {Eyink}},\ }\href@noop {} {\bibfield  {journal} {\bibinfo  {journal} {Journal
  of Turbulence}\ }\textbf {\bibinfo {volume} {9}},\ \bibinfo {pages} {N31}
  (\bibinfo {year} {2008})}\BibitemShut {NoStop}%
\end{thebibliography}%


\begin{thebibliography}{45}%
\makeatletter
\providecommand \@ifxundefined [1]{%
 \@ifx{#1\undefined}
}%
\providecommand \@ifnum [1]{%
 \ifnum #1\expandafter \@firstoftwo
 \else \expandafter \@secondoftwo
 \fi
}%
\providecommand \@ifx [1]{%
 \ifx #1\expandafter \@firstoftwo
 \else \expandafter \@secondoftwo
 \fi
}%
\providecommand \natexlab [1]{#1}%
\providecommand \enquote  [1]{``#1''}%
\providecommand \bibnamefont  [1]{#1}%
\providecommand \bibfnamefont [1]{#1}%
\providecommand \citenamefont [1]{#1}%
\providecommand \href@noop [0]{\@secondoftwo}%
\providecommand \href [0]{\begingroup \@sanitize@url \@href}%
\providecommand \@href[1]{\@@startlink{#1}\@@href}%
\providecommand \@@href[1]{\endgroup#1\@@endlink}%
\providecommand \@sanitize@url [0]{\catcode `\\12\catcode `\$12\catcode
  `\&12\catcode `\#12\catcode `\^12\catcode `\_12\catcode `\%12\relax}%
\providecommand \@@startlink[1]{}%
\providecommand \@@endlink[0]{}%
\providecommand \url  [0]{\begingroup\@sanitize@url \@url }%
\providecommand \@url [1]{\endgroup\@href {#1}{\urlprefix }}%
\providecommand \urlprefix  [0]{URL }%
\providecommand \Eprint [0]{\href }%
\providecommand \doibase [0]{http://dx.doi.org/}%
\providecommand \selectlanguage [0]{\@gobble}%
\providecommand \bibinfo  [0]{\@secondoftwo}%
\providecommand \bibfield  [0]{\@secondoftwo}%
\providecommand \translation [1]{[#1]}%
\providecommand \BibitemOpen [0]{}%
\providecommand \bibitemStop [0]{}%
\providecommand \bibitemNoStop [0]{.\EOS\space}%
\providecommand \EOS [0]{\spacefactor3000\relax}%
\providecommand \BibitemShut  [1]{\csname bibitem#1\endcsname}%
\let\auto@bib@innerbib\@empty
\bibitem [{\citenamefont {Dimotakis}(2005)}]{Dimotakis:2005aa}%
  \BibitemOpen
  \bibfield  {author} {\bibinfo {author} {\bibfnamefont {P.~E.}\ \bibnamefont
  {Dimotakis}},\ }\href {\doibase
  https://doi.org/10.1146/annurev.fluid.36.050802.122015} {\bibfield  {journal}
  {\bibinfo  {journal} {Annual Review of Fluid Mechanics}\ }\textbf {\bibinfo
  {volume} {37}},\ \bibinfo {pages} {329} (\bibinfo {year} {2005})}\BibitemShut
  {NoStop}%
\bibitem [{\citenamefont {Haller}(2015)}]{Haller:2015aa}%
  \BibitemOpen
  \bibfield  {author} {\bibinfo {author} {\bibfnamefont {G.}~\bibnamefont
  {Haller}},\ }\href {\doibase
  https://doi.org/10.1146/annurev-fluid-010313-141322} {\bibfield  {journal}
  {\bibinfo  {journal} {Annual Review of Fluid Mechanics}\ }\textbf {\bibinfo
  {volume} {47}},\ \bibinfo {pages} {137} (\bibinfo {year} {2015})}\BibitemShut
  {NoStop}%
\bibitem [{\citenamefont {Villermaux}(2019)}]{Villermaux:2019aa}%
  \BibitemOpen
  \bibfield  {author} {\bibinfo {author} {\bibfnamefont {E.}~\bibnamefont
  {Villermaux}},\ }\href@noop {} {\bibfield  {journal} {\bibinfo  {journal}
  {Annual Review of Fluid Mechanics}\ }\textbf {\bibinfo {volume} {51}},\
  \bibinfo {pages} {245} (\bibinfo {year} {2019})}\BibitemShut {NoStop}%
\bibitem [{\citenamefont {Rivlin}\ and\ \citenamefont
  {Sawyers}(1971)}]{Rivlin:1971aa}%
  \BibitemOpen
  \bibfield  {author} {\bibinfo {author} {\bibfnamefont {R.~S.}\ \bibnamefont
  {Rivlin}}\ and\ \bibinfo {author} {\bibfnamefont {K.~N.}\ \bibnamefont
  {Sawyers}},\ }\href {\doibase
  https://doi.org/10.1146/annurev.fl.03.010171.001001} {\bibfield  {journal}
  {\bibinfo  {journal} {Annual Review of Fluid Mechanics}\ }\textbf {\bibinfo
  {volume} {3}},\ \bibinfo {pages} {117} (\bibinfo {year} {1971})}\BibitemShut
  {NoStop}%
\bibitem [{\citenamefont {Stone}(1994)}]{Stone:1994aa}%
  \BibitemOpen
  \bibfield  {author} {\bibinfo {author} {\bibfnamefont {H.~A.}\ \bibnamefont
  {Stone}},\ }\href {\doibase
  https://doi.org/10.1146/annurev.fl.26.010194.000433} {\bibfield  {journal}
  {\bibinfo  {journal} {Annual Review of Fluid Mechanics}\ }\textbf {\bibinfo
  {volume} {26}},\ \bibinfo {pages} {65} (\bibinfo {year} {1994})}\BibitemShut
  {NoStop}%
\bibitem [{\citenamefont {Voth}\ and\ \citenamefont
  {Soldati}(2017)}]{Voth:2017aa}%
  \BibitemOpen
  \bibfield  {author} {\bibinfo {author} {\bibfnamefont {G.~A.}\ \bibnamefont
  {Voth}}\ and\ \bibinfo {author} {\bibfnamefont {A.}~\bibnamefont {Soldati}},\
  }\href {\doibase https://doi.org/10.1146/annurev-fluid-010816-060135}
  {\bibfield  {journal} {\bibinfo  {journal} {Annual Review of Fluid
  Mechanics}\ }\textbf {\bibinfo {volume} {49}},\ \bibinfo {pages} {249}
  (\bibinfo {year} {2017})}\BibitemShut {NoStop}%
\bibitem [{\citenamefont {Libby}\ and\ \citenamefont
  {Williams}(1976)}]{Libby:1976aa}%
  \BibitemOpen
  \bibfield  {author} {\bibinfo {author} {\bibfnamefont {P.~A.}\ \bibnamefont
  {Libby}}\ and\ \bibinfo {author} {\bibfnamefont {F.~A.}\ \bibnamefont
  {Williams}},\ }\href {\doibase
  https://doi.org/10.1146/annurev.fl.08.010176.002031} {\bibfield  {journal}
  {\bibinfo  {journal} {Annual Review of Fluid Mechanics}\ }\textbf {\bibinfo
  {volume} {8}},\ \bibinfo {pages} {351} (\bibinfo {year} {1976})}\BibitemShut
  {NoStop}%
\bibitem [{\citenamefont {T{\'e}l}\ \emph {et~al.}(2005)\citenamefont
  {T{\'e}l}, \citenamefont {de~Moura}, \citenamefont {Grebogi},\ and\
  \citenamefont {K{\'a}rolyi}}]{Tel:2005aa}%
  \BibitemOpen
  \bibfield  {author} {\bibinfo {author} {\bibfnamefont {T.}~\bibnamefont
  {T{\'e}l}}, \bibinfo {author} {\bibfnamefont {A.}~\bibnamefont {de~Moura}},
  \bibinfo {author} {\bibfnamefont {C.}~\bibnamefont {Grebogi}}, \ and\
  \bibinfo {author} {\bibfnamefont {G.}~\bibnamefont {K{\'a}rolyi}},\
  }\href@noop {} {\bibfield  {journal} {\bibinfo  {journal} {Physics Reports}\
  }\textbf {\bibinfo {volume} {413}},\ \bibinfo {pages} {91 } (\bibinfo {year}
  {2005})}\BibitemShut {NoStop}%
\bibitem [{\citenamefont {Aref}\ \emph {et~al.}(2017)\citenamefont {Aref},
  \citenamefont {Blake}, \citenamefont {Budi{\v{s}}i{\'c}}, \citenamefont
  {Cardoso}, \citenamefont {Cartwright}, \citenamefont {Clercx}, \citenamefont
  {El~Omari}, \citenamefont {Feudel}, \citenamefont {Golestanian},
  \citenamefont {Gouillart} \emph {et~al.}}]{Aref:2017aa}%
  \BibitemOpen
  \bibfield  {author} {\bibinfo {author} {\bibfnamefont {H.}~\bibnamefont
  {Aref}}, \bibinfo {author} {\bibfnamefont {J.~R.}\ \bibnamefont {Blake}},
  \bibinfo {author} {\bibfnamefont {M.}~\bibnamefont {Budi{\v{s}}i{\'c}}},
  \bibinfo {author} {\bibfnamefont {S.~S.}\ \bibnamefont {Cardoso}}, \bibinfo
  {author} {\bibfnamefont {J.~H.}\ \bibnamefont {Cartwright}}, \bibinfo
  {author} {\bibfnamefont {H.~J.}\ \bibnamefont {Clercx}}, \bibinfo {author}
  {\bibfnamefont {K.}~\bibnamefont {El~Omari}}, \bibinfo {author}
  {\bibfnamefont {U.}~\bibnamefont {Feudel}}, \bibinfo {author} {\bibfnamefont
  {R.}~\bibnamefont {Golestanian}}, \bibinfo {author} {\bibfnamefont
  {E.}~\bibnamefont {Gouillart}},  \emph {et~al.},\ }\href@noop {} {\bibfield
  {journal} {\bibinfo  {journal} {Reviews of Modern Physics}\ }\textbf
  {\bibinfo {volume} {89}},\ \bibinfo {pages} {025007} (\bibinfo {year}
  {2017})}\BibitemShut {NoStop}%
\bibitem [{\citenamefont {Sreenivasan}(2019)}]{Sreenivasan:2019aa}%
  \BibitemOpen
  \bibfield  {author} {\bibinfo {author} {\bibfnamefont {K.~R.}\ \bibnamefont
  {Sreenivasan}},\ }\href@noop {} {\bibfield  {journal} {\bibinfo  {journal}
  {Proceedings of the National Academy of Sciences}\ }\textbf {\bibinfo
  {volume} {116}},\ \bibinfo {pages} {18175} (\bibinfo {year}
  {2019})}\BibitemShut {NoStop}%
\bibitem [{\citenamefont {Villermaux}\ and\ \citenamefont
  {Duplat}(2003)}]{Villermaux:2003ab}%
  \BibitemOpen
  \bibfield  {author} {\bibinfo {author} {\bibfnamefont {E.}~\bibnamefont
  {Villermaux}}\ and\ \bibinfo {author} {\bibfnamefont {J.}~\bibnamefont
  {Duplat}},\ }\href {\doibase 10.1103/PhysRevLett.91.184501} {\bibfield
  {journal} {\bibinfo  {journal} {Phys. Rev. Lett.}\ }\textbf {\bibinfo
  {volume} {91}},\ \bibinfo {pages} {184501} (\bibinfo {year}
  {2003})}\BibitemShut {NoStop}%
\bibitem [{\citenamefont {Haller}\ and\ \citenamefont
  {Sapsis}(2008)}]{Haller:2008aa}%
  \BibitemOpen
  \bibfield  {author} {\bibinfo {author} {\bibfnamefont {G.}~\bibnamefont
  {Haller}}\ and\ \bibinfo {author} {\bibfnamefont {T.}~\bibnamefont
  {Sapsis}},\ }\href {\doibase http://dx.doi.org/10.1016/j.physd.2007.09.027}
  {\bibfield  {journal} {\bibinfo  {journal} {Physica D: Nonlinear Phenomena}\
  }\textbf {\bibinfo {volume} {237}},\ \bibinfo {pages} {573 } (\bibinfo {year}
  {2008})}\BibitemShut {NoStop}%
\bibitem [{\citenamefont {Cocke}(1969)}]{Cocke:1969aa}%
  \BibitemOpen
  \bibfield  {author} {\bibinfo {author} {\bibfnamefont {W.~J.}\ \bibnamefont
  {Cocke}},\ }\href@noop {} {\bibfield  {journal} {\bibinfo  {journal} {The
  Physics of Fluids}\ }\textbf {\bibinfo {volume} {12}},\ \bibinfo {pages}
  {2488} (\bibinfo {year} {1969})}\BibitemShut {NoStop}%
\bibitem [{\citenamefont {Girimaji}\ and\ \citenamefont
  {Pope}(1990)}]{Girimaji:1990aa}%
  \BibitemOpen
  \bibfield  {author} {\bibinfo {author} {\bibfnamefont {S.~S.}\ \bibnamefont
  {Girimaji}}\ and\ \bibinfo {author} {\bibfnamefont {S.~B.}\ \bibnamefont
  {Pope}},\ }\href {\doibase 10.1017/S0022112090003330} {\bibfield  {journal}
  {\bibinfo  {journal} {Journal of Fluid Mechanics}\ }\textbf {\bibinfo
  {volume} {220}},\ \bibinfo {pages} {427} (\bibinfo {year}
  {1990})}\BibitemShut {NoStop}%
\bibitem [{\citenamefont {Duplat}, \citenamefont {Innocenti},\ and\
  \citenamefont {Villermaux}(2010)}]{Duplat:2010ab}%
  \BibitemOpen
  \bibfield  {author} {\bibinfo {author} {\bibfnamefont {J.}~\bibnamefont
  {Duplat}}, \bibinfo {author} {\bibfnamefont {C.}~\bibnamefont {Innocenti}}, \
  and\ \bibinfo {author} {\bibfnamefont {E.}~\bibnamefont {Villermaux}},\
  }\href@noop {} {\bibfield  {journal} {\bibinfo  {journal} {Phys. Fluids}\
  }\textbf {\bibinfo {volume} {22}},\ \bibinfo {pages} {035104} (\bibinfo
  {year} {2010})}\BibitemShut {NoStop}%
\bibitem [{\citenamefont {Kalda}(2000)}]{Kalda:2000aa}%
  \BibitemOpen
  \bibfield  {author} {\bibinfo {author} {\bibfnamefont {J.}~\bibnamefont
  {Kalda}},\ }\href {\doibase 10.1103/PhysRevLett.84.471} {\bibfield  {journal}
  {\bibinfo  {journal} {Phys. Rev. Lett.}\ }\textbf {\bibinfo {volume} {84}},\
  \bibinfo {pages} {471} (\bibinfo {year} {2000})}\BibitemShut {NoStop}%
\bibitem [{\citenamefont {Matsuoka}\ and\ \citenamefont
  {Hiraide}(2015)}]{Matsuoka:2015aa}%
  \BibitemOpen
  \bibfield  {author} {\bibinfo {author} {\bibfnamefont {C.}~\bibnamefont
  {Matsuoka}}\ and\ \bibinfo {author} {\bibfnamefont {K.}~\bibnamefont
  {Hiraide}},\ }\href@noop {} {\bibfield  {journal} {\bibinfo  {journal}
  {Chaos: An Interdisciplinary Journal of Nonlinear Science}\ }\textbf
  {\bibinfo {volume} {25}},\ \bibinfo {pages} {103110} (\bibinfo {year}
  {2015})}\BibitemShut {NoStop}%
\bibitem [{\citenamefont {Newhouse}\ and\ \citenamefont
  {Pignataro}(1993)}]{Newhouse:1993aa}%
  \BibitemOpen
  \bibfield  {author} {\bibinfo {author} {\bibfnamefont {S.}~\bibnamefont
  {Newhouse}}\ and\ \bibinfo {author} {\bibfnamefont {T.}~\bibnamefont
  {Pignataro}},\ }\href {\doibase 10.1007/BF01048189} {\bibfield  {journal}
  {\bibinfo  {journal} {Journal of Statistical Physics}\ }\textbf {\bibinfo
  {volume} {72}},\ \bibinfo {pages} {1331} (\bibinfo {year}
  {1993})}\BibitemShut {NoStop}%
\bibitem [{\citenamefont {Goto}\ and\ \citenamefont
  {Kida}(2002)}]{Goto:2002aa}%
  \BibitemOpen
  \bibfield  {author} {\bibinfo {author} {\bibfnamefont {S.}~\bibnamefont
  {Goto}}\ and\ \bibinfo {author} {\bibfnamefont {S.}~\bibnamefont {Kida}},\
  }\href {\doibase 10.1088/1468-5248/3/1/017} {\bibfield  {journal} {\bibinfo
  {journal} {Journal of Turbulence}\ }\textbf {\bibinfo {volume} {3}},\
  \bibinfo {pages} {N17} (\bibinfo {year} {2002})}\BibitemShut {NoStop}%
\bibitem [{\citenamefont {Li}\ \emph {et~al.}(2008)\citenamefont {Li},
  \citenamefont {Perlman}, \citenamefont {Wan}, \citenamefont {Yang},
  \citenamefont {Meneveau}, \citenamefont {Burns}, \citenamefont {Chen},
  \citenamefont {Szalay},\ and\ \citenamefont {Eyink}}]{Li:2008aa}%
  \BibitemOpen
  \bibfield  {author} {\bibinfo {author} {\bibfnamefont {Y.}~\bibnamefont
  {Li}}, \bibinfo {author} {\bibfnamefont {E.}~\bibnamefont {Perlman}},
  \bibinfo {author} {\bibfnamefont {M.}~\bibnamefont {Wan}}, \bibinfo {author}
  {\bibfnamefont {Y.}~\bibnamefont {Yang}}, \bibinfo {author} {\bibfnamefont
  {C.}~\bibnamefont {Meneveau}}, \bibinfo {author} {\bibfnamefont
  {R.}~\bibnamefont {Burns}}, \bibinfo {author} {\bibfnamefont
  {S.}~\bibnamefont {Chen}}, \bibinfo {author} {\bibfnamefont {A.}~\bibnamefont
  {Szalay}}, \ and\ \bibinfo {author} {\bibfnamefont {G.}~\bibnamefont
  {Eyink}},\ }\href@noop {} {\bibfield  {journal} {\bibinfo  {journal} {Journal
  of Turbulence}\ }\textbf {\bibinfo {volume} {9}},\ \bibinfo {pages} {N31}
  (\bibinfo {year} {2008})}\BibitemShut {NoStop}%
\bibitem [{\citenamefont {Yeung}(2002)}]{Yeung:2002aa}%
  \BibitemOpen
  \bibfield  {author} {\bibinfo {author} {\bibfnamefont {P.~K.}\ \bibnamefont
  {Yeung}},\ }\href {\doibase
  https://doi.org/10.1146/annurev.fluid.34.082101.170725} {\bibfield  {journal}
  {\bibinfo  {journal} {Annual Review of Fluid Mechanics}\ }\textbf {\bibinfo
  {volume} {34}},\ \bibinfo {pages} {115} (\bibinfo {year} {2002})}\BibitemShut
  {NoStop}%
\bibitem [{\citenamefont {Lester}\ and\ \citenamefont
  {Dentz}(2025)}]{Lester:2025aa}%
  \BibitemOpen
  \bibfield  {author} {\bibinfo {author} {\bibfnamefont {D.~R.}\ \bibnamefont
  {Lester}}\ and\ \bibinfo {author} {\bibfnamefont {M.}~\bibnamefont {Dentz}},\
  }\href@noop {} {\bibfield  {journal} {\bibinfo  {journal} {arXiv}\ }\textbf
  {\bibinfo {volume} {submit/6850609}} (\bibinfo {year} {2025})}\BibitemShut
  {NoStop}%
\bibitem [{sup()}]{suppmat}%
  \BibitemOpen
  \href@noop {} {\enquote {\bibinfo {title} {Supplementary material},}\
  }\BibitemShut {NoStop}%
\bibitem [{\citenamefont {Meneveau}\ and\ \citenamefont
  {Sreenivasan}(1991)}]{Meneveau:1991aa}%
  \BibitemOpen
  \bibfield  {author} {\bibinfo {author} {\bibfnamefont {C.}~\bibnamefont
  {Meneveau}}\ and\ \bibinfo {author} {\bibfnamefont {K.~R.}\ \bibnamefont
  {Sreenivasan}},\ }\href {\doibase 10.1017/S0022112091001830} {\bibfield
  {journal} {\bibinfo  {journal} {Journal of Fluid Mechanics}\ }\textbf
  {\bibinfo {volume} {224}},\ \bibinfo {pages} {429–484} (\bibinfo {year}
  {1991})}\BibitemShut {NoStop}%
\bibitem [{\citenamefont {Lester}\ \emph {et~al.}(2025)\citenamefont {Lester},
  \citenamefont {Metcalfe}, \citenamefont {Trefry},\ and\ \citenamefont
  {Dentz}}]{Lester:2025ac}%
  \BibitemOpen
  \bibfield  {author} {\bibinfo {author} {\bibfnamefont {D.~R.}\ \bibnamefont
  {Lester}}, \bibinfo {author} {\bibfnamefont {G.}~\bibnamefont {Metcalfe}},
  \bibinfo {author} {\bibfnamefont {M.}~\bibnamefont {Trefry}}, \ and\ \bibinfo
  {author} {\bibfnamefont {M.}~\bibnamefont {Dentz}},\ }\href {\doibase
  10.1017/jfm.2025.10551} {\bibfield  {journal} {\bibinfo  {journal} {Journal
  of Fluid Mechanics}\ }\textbf {\bibinfo {volume} {1018}},\ \bibinfo {pages}
  {A35} (\bibinfo {year} {2025})}\BibitemShut {NoStop}%
\bibitem [{\citenamefont {Dresselhaus}\ and\ \citenamefont
  {Tabor}(1992)}]{Dresselhaus:1992aa}%
  \BibitemOpen
  \bibfield  {author} {\bibinfo {author} {\bibfnamefont {E.}~\bibnamefont
  {Dresselhaus}}\ and\ \bibinfo {author} {\bibfnamefont {M.}~\bibnamefont
  {Tabor}},\ }\href {\doibase 10.1017/S0022112092001460} {\bibfield  {journal}
  {\bibinfo  {journal} {Journal of Fluid Mechanics}\ }\textbf {\bibinfo
  {volume} {236}},\ \bibinfo {pages} {415} (\bibinfo {year}
  {1992})}\BibitemShut {NoStop}%
\bibitem [{\citenamefont {Tabor}\ and\ \citenamefont
  {Klapper}(1994)}]{Tabor:1994aa}%
  \BibitemOpen
  \bibfield  {author} {\bibinfo {author} {\bibfnamefont {M.}~\bibnamefont
  {Tabor}}\ and\ \bibinfo {author} {\bibfnamefont {I.}~\bibnamefont
  {Klapper}},\ }\href {\doibase https://doi.org/10.1016/0960-0779(94)90137-6}
  {\bibfield  {journal} {\bibinfo  {journal} {Chaos, Solitons and Fractals}\
  }\textbf {\bibinfo {volume} {4}},\ \bibinfo {pages} {1031} (\bibinfo {year}
  {1994})}\BibitemShut {NoStop}%
\bibitem [{\citenamefont {Lester}\ \emph {et~al.}(2018)\citenamefont {Lester},
  \citenamefont {Dentz}, \citenamefont {Borgne},\ and\ \citenamefont
  {Barros}}]{Lester:2018aa}%
  \BibitemOpen
  \bibfield  {author} {\bibinfo {author} {\bibfnamefont {D.~R.}\ \bibnamefont
  {Lester}}, \bibinfo {author} {\bibfnamefont {M.}~\bibnamefont {Dentz}},
  \bibinfo {author} {\bibfnamefont {T.~L.}\ \bibnamefont {Borgne}}, \ and\
  \bibinfo {author} {\bibfnamefont {F.~P. J.~D.}\ \bibnamefont {Barros}},\
  }\href@noop {} {\bibfield  {journal} {\bibinfo  {journal} {Journal of Fluid
  Mechanics}\ }\textbf {\bibinfo {volume} {855}},\ \bibinfo {pages} {770}
  (\bibinfo {year} {2018})}\BibitemShut {NoStop}%
\bibitem [{\citenamefont {Dentz}\ \emph {et~al.}(2016)\citenamefont {Dentz},
  \citenamefont {Lester}, \citenamefont {Le~Borgne},\ and\ \citenamefont
  {de~Barros}}]{Dentz:2016aa}%
  \BibitemOpen
  \bibfield  {author} {\bibinfo {author} {\bibfnamefont {M.}~\bibnamefont
  {Dentz}}, \bibinfo {author} {\bibfnamefont {D.~R.}\ \bibnamefont {Lester}},
  \bibinfo {author} {\bibfnamefont {T.}~\bibnamefont {Le~Borgne}}, \ and\
  \bibinfo {author} {\bibfnamefont {F.~P.~J.}\ \bibnamefont {de~Barros}},\
  }\href {\doibase 10.1103/PhysRevE.94.061102} {\bibfield  {journal} {\bibinfo
  {journal} {Phys. Rev. E}\ }\textbf {\bibinfo {volume} {94}},\ \bibinfo
  {pages} {061102} (\bibinfo {year} {2016})}\BibitemShut {NoStop}%
\bibitem [{\citenamefont {Lester}\ \emph {et~al.}(2022)\citenamefont {Lester},
  \citenamefont {Dentz}, \citenamefont {Bandopadhyay},\ and\ \citenamefont
  {Borgne}}]{Lester:2022aa}%
  \BibitemOpen
  \bibfield  {author} {\bibinfo {author} {\bibfnamefont {D.~R.}\ \bibnamefont
  {Lester}}, \bibinfo {author} {\bibfnamefont {M.}~\bibnamefont {Dentz}},
  \bibinfo {author} {\bibfnamefont {A.}~\bibnamefont {Bandopadhyay}}, \ and\
  \bibinfo {author} {\bibfnamefont {T.~L.}\ \bibnamefont {Borgne}},\
  }\href@noop {} {\bibfield  {journal} {\bibinfo  {journal} {Journal of Fluid
  Mechanics}\ }\textbf {\bibinfo {volume} {945}},\ \bibinfo {pages} {A18}
  (\bibinfo {year} {2022})}\BibitemShut {NoStop}%
\bibitem [{\citenamefont {Kree}\ and\ \citenamefont
  {Villermaux}(2017)}]{Kree:2017aa}%
  \BibitemOpen
  \bibfield  {author} {\bibinfo {author} {\bibfnamefont {M.}~\bibnamefont
  {Kree}}\ and\ \bibinfo {author} {\bibfnamefont {E.}~\bibnamefont
  {Villermaux}},\ }\href {\doibase 10.1103/PhysRevFluids.2.104502} {\bibfield
  {journal} {\bibinfo  {journal} {Phys. Rev. Fluids}\ }\textbf {\bibinfo
  {volume} {2}},\ \bibinfo {pages} {104502} (\bibinfo {year}
  {2017})}\BibitemShut {NoStop}%
\bibitem [{\citenamefont {Souzy}\ \emph {et~al.}(2017)\citenamefont {Souzy},
  \citenamefont {Lhuissier}, \citenamefont {Villermaux},\ and\ \citenamefont
  {Metzger}}]{Souzy:2017aa}%
  \BibitemOpen
  \bibfield  {author} {\bibinfo {author} {\bibfnamefont {M.}~\bibnamefont
  {Souzy}}, \bibinfo {author} {\bibfnamefont {H.}~\bibnamefont {Lhuissier}},
  \bibinfo {author} {\bibfnamefont {E.}~\bibnamefont {Villermaux}}, \ and\
  \bibinfo {author} {\bibfnamefont {B.}~\bibnamefont {Metzger}},\ }\href@noop
  {} {\bibfield  {journal} {\bibinfo  {journal} {Journal of Fluid Mechanics}\
  }\textbf {\bibinfo {volume} {812}},\ \bibinfo {pages} {611} (\bibinfo {year}
  {2017})}\BibitemShut {NoStop}%
\bibitem [{\citenamefont {Villermaux}\ and\ \citenamefont
  {Gagne}(1994)}]{Villermaux:1994aa}%
  \BibitemOpen
  \bibfield  {author} {\bibinfo {author} {\bibfnamefont {E.}~\bibnamefont
  {Villermaux}}\ and\ \bibinfo {author} {\bibfnamefont {Y.}~\bibnamefont
  {Gagne}},\ }\href {\doibase 10.1103/PhysRevLett.73.252} {\bibfield  {journal}
  {\bibinfo  {journal} {Phys. Rev. Lett.}\ }\textbf {\bibinfo {volume} {73}},\
  \bibinfo {pages} {252} (\bibinfo {year} {1994})}\BibitemShut {NoStop}%
\bibitem [{\citenamefont {Voth}, \citenamefont {Haller},\ and\ \citenamefont
  {Gollub}(2002)}]{Voth:2002aa}%
  \BibitemOpen
  \bibfield  {author} {\bibinfo {author} {\bibfnamefont {G.~A.}\ \bibnamefont
  {Voth}}, \bibinfo {author} {\bibfnamefont {G.}~\bibnamefont {Haller}}, \ and\
  \bibinfo {author} {\bibfnamefont {J.~P.}\ \bibnamefont {Gollub}},\ }\href
  {\doibase 10.1103/PhysRevLett.88.254501} {\bibfield  {journal} {\bibinfo
  {journal} {Phys. Rev. Lett.}\ }\textbf {\bibinfo {volume} {88}},\ \bibinfo
  {pages} {254501} (\bibinfo {year} {2002})}\BibitemShut {NoStop}%
\bibitem [{\citenamefont {Heyman}\ \emph {et~al.}(2020)\citenamefont {Heyman},
  \citenamefont {Lester}, \citenamefont {Turuban}, \citenamefont
  {M{\'e}heust},\ and\ \citenamefont {Le~Borgne}}]{Heyman:2020aa}%
  \BibitemOpen
  \bibfield  {author} {\bibinfo {author} {\bibfnamefont {J.}~\bibnamefont
  {Heyman}}, \bibinfo {author} {\bibfnamefont {D.~R.}\ \bibnamefont {Lester}},
  \bibinfo {author} {\bibfnamefont {R.}~\bibnamefont {Turuban}}, \bibinfo
  {author} {\bibfnamefont {Y.}~\bibnamefont {M{\'e}heust}}, \ and\ \bibinfo
  {author} {\bibfnamefont {T.}~\bibnamefont {Le~Borgne}},\ }\href {\doibase
  10.1073/pnas.2002858117} {\bibfield  {journal} {\bibinfo  {journal}
  {Proceedings of the National Academy of Sciences}\ }\textbf {\bibinfo
  {volume} {117}},\ \bibinfo {pages} {13359} (\bibinfo {year}
  {2020})}\BibitemShut {NoStop}%
\bibitem [{\citenamefont {Le~Borgne}, \citenamefont {Dentz},\ and\
  \citenamefont {Carrera}(2008)}]{Le-Borgne:2008aa}%
  \BibitemOpen
  \bibfield  {author} {\bibinfo {author} {\bibfnamefont {T.}~\bibnamefont
  {Le~Borgne}}, \bibinfo {author} {\bibfnamefont {M.}~\bibnamefont {Dentz}}, \
  and\ \bibinfo {author} {\bibfnamefont {J.}~\bibnamefont {Carrera}},\ }\href
  {\doibase 10.1103/PhysRevLett.101.090601} {\bibfield  {journal} {\bibinfo
  {journal} {Phys. Rev. Lett.}\ }\textbf {\bibinfo {volume} {101}},\ \bibinfo
  {pages} {090601} (\bibinfo {year} {2008})}\BibitemShut {NoStop}%
\bibitem [{\citenamefont {Lumley}(2007)}]{lumley:2007aa}%
  \BibitemOpen
  \bibfield  {author} {\bibinfo {author} {\bibfnamefont {J.~L.}\ \bibnamefont
  {Lumley}},\ }\href@noop {} {\emph {\bibinfo {title} {Stochastic tools in
  turbulence}}}\ (\bibinfo  {publisher} {Dover Publications},\ \bibinfo
  {address} {Mineola, New York},\ \bibinfo {year} {2007})\BibitemShut {NoStop}%
\bibitem [{\citenamefont {Pope}(1994)}]{Pope:1994aa}%
  \BibitemOpen
  \bibfield  {author} {\bibinfo {author} {\bibfnamefont {S.~B.}\ \bibnamefont
  {Pope}},\ }\href {\doibase
  https://doi.org/10.1146/annurev.fl.26.010194.000323} {\bibfield  {journal}
  {\bibinfo  {journal} {Annual Review of Fluid Mechanics}\ }\textbf {\bibinfo
  {volume} {26}},\ \bibinfo {pages} {23} (\bibinfo {year} {1994})}\BibitemShut
  {NoStop}%
\bibitem [{\citenamefont {Falkovich}, \citenamefont
  {Gaw\ifmmode~\mbox{\c{e}}\else \c{e}\fi{}dzki},\ and\ \citenamefont
  {Vergassola}(2001)}]{Falkovich:2001ab}%
  \BibitemOpen
  \bibfield  {author} {\bibinfo {author} {\bibfnamefont {G.}~\bibnamefont
  {Falkovich}}, \bibinfo {author} {\bibfnamefont {K.}~\bibnamefont
  {Gaw\ifmmode~\mbox{\c{e}}\else \c{e}\fi{}dzki}}, \ and\ \bibinfo {author}
  {\bibfnamefont {M.}~\bibnamefont {Vergassola}},\ }\href@noop {} {\bibfield
  {journal} {\bibinfo  {journal} {Rev. Mod. Phys.}\ }\textbf {\bibinfo {volume}
  {73}},\ \bibinfo {pages} {913} (\bibinfo {year} {2001})}\BibitemShut
  {NoStop}%
\bibitem [{\citenamefont {Yu}\ \emph {et~al.}(2023)\citenamefont {Yu},
  \citenamefont {Fouxon}, \citenamefont {Wang}, \citenamefont {Li},
  \citenamefont {Yuan}, \citenamefont {Mao},\ and\ \citenamefont
  {Mond}}]{Yu:2023aa}%
  \BibitemOpen
  \bibfield  {author} {\bibinfo {author} {\bibfnamefont {H.}~\bibnamefont
  {Yu}}, \bibinfo {author} {\bibfnamefont {I.}~\bibnamefont {Fouxon}}, \bibinfo
  {author} {\bibfnamefont {J.}~\bibnamefont {Wang}}, \bibinfo {author}
  {\bibfnamefont {X.}~\bibnamefont {Li}}, \bibinfo {author} {\bibfnamefont
  {L.}~\bibnamefont {Yuan}}, \bibinfo {author} {\bibfnamefont {S.}~\bibnamefont
  {Mao}}, \ and\ \bibinfo {author} {\bibfnamefont {M.}~\bibnamefont {Mond}},\
  }\href@noop {} {\bibfield  {journal} {\bibinfo  {journal} {Physics of
  Fluids}\ }\textbf {\bibinfo {volume} {35}},\ \bibinfo {pages} {125114}
  (\bibinfo {year} {2023})}\BibitemShut {NoStop}%
\bibitem [{\citenamefont {Bresler}\ \emph {et~al.}(1997)\citenamefont
  {Bresler}, \citenamefont {Shinbrot}, \citenamefont {Metcalfe},\ and\
  \citenamefont {Ottino}}]{Bresler:1997aa}%
  \BibitemOpen
  \bibfield  {author} {\bibinfo {author} {\bibfnamefont {L.}~\bibnamefont
  {Bresler}}, \bibinfo {author} {\bibfnamefont {T.}~\bibnamefont {Shinbrot}},
  \bibinfo {author} {\bibfnamefont {G.}~\bibnamefont {Metcalfe}}, \ and\
  \bibinfo {author} {\bibfnamefont {J.~M.}\ \bibnamefont {Ottino}},\
  }\href@noop {} {\bibfield  {journal} {\bibinfo  {journal} {Chemical
  Engineering Science}\ }\textbf {\bibinfo {volume} {52}},\ \bibinfo {pages}
  {1623} (\bibinfo {year} {1997})}\BibitemShut {NoStop}%
\bibitem [{\citenamefont {Kraichnan}(1970)}]{Kraichnan:1970aa}%
  \BibitemOpen
  \bibfield  {author} {\bibinfo {author} {\bibfnamefont {R.~H.}\ \bibnamefont
  {Kraichnan}},\ }\href {\doibase http://dx.doi.org/10.1063/1.1692799}
  {\bibfield  {journal} {\bibinfo  {journal} {Physics of Fluids}\ }\textbf
  {\bibinfo {volume} {13}},\ \bibinfo {pages} {22} (\bibinfo {year}
  {1970})}\BibitemShut {NoStop}%
\bibitem [{\citenamefont {David}(2011)}]{David:2011aa}%
  \BibitemOpen
  \bibfield  {author} {\bibinfo {author} {\bibfnamefont {H.~A.}\ \bibnamefont
  {David}},\ }\enquote {\bibinfo {title} {Order statistics},}\ \ (\bibinfo
  {publisher} {Springer Berlin Heidelberg},\ \bibinfo {address} {Berlin,
  Heidelberg},\ \bibinfo {year} {2011})\BibitemShut {NoStop}%
\bibitem [{\citenamefont {Kolmogorov}(1991)}]{Kolmogorov:1991aa}%
  \BibitemOpen
  \bibfield  {author} {\bibinfo {author} {\bibfnamefont {A.~N.}\ \bibnamefont
  {Kolmogorov}},\ }\href {http://www.jstor.org/stable/51980} {\bibfield
  {journal} {\bibinfo  {journal} {Proceedings: Mathematical and Physical
  Sciences}\ }\textbf {\bibinfo {volume} {434}},\ \bibinfo {pages} {9}
  (\bibinfo {year} {1991})}\BibitemShut {NoStop}%
\bibitem [{\citenamefont {Ho}, \citenamefont {Armua},\ and\ \citenamefont
  {Berera}(2020)}]{Ho:2020aa}%
  \BibitemOpen
  \bibfield  {author} {\bibinfo {author} {\bibfnamefont {R.~D. J.~G.}\
  \bibnamefont {Ho}}, \bibinfo {author} {\bibfnamefont {A.}~\bibnamefont
  {Armua}}, \ and\ \bibinfo {author} {\bibfnamefont {A.}~\bibnamefont
  {Berera}},\ }\href {\doibase 10.1103/PhysRevFluids.5.024602} {\bibfield
  {journal} {\bibinfo  {journal} {Phys. Rev. Fluids}\ }\textbf {\bibinfo
  {volume} {5}},\ \bibinfo {pages} {024602} (\bibinfo {year}
  {2020})}\BibitemShut {NoStop}%
\end{thebibliography}%

\end{document}


\preprint{APS/123-QED}

\title{Supplementary Material for ``Line Stretching in Random Flows''}

\author{D. R. Lester}
\affiliation{School of Engineering, RMIT University, 3000 Melbourne, Australia}
\email{daniel.lester@rmit.edu.au}
\author{M. Dentz}
\affiliation{Spanish National Research Council (IDAEA-CSIC), 08034 Barcelona, Spain}
\date{\today}

\pacs{}

\maketitle

\renewcommand{\theequation}{S.\arabic{equation}}

\section{Numerical Simulations}

\subsection{Homogeneous isotropic turbulence} 

In Fig.1b of the main paper we demonstrate evolution of the finite-time topological entropy (FTTE) $h(t)$ for the case of forced homogeneous isotropic turbulence based on direct numerical simulation (DNS) data (Johns Hopkins Turbulence Database) at Taylor-scale Reynolds number $Re_\lambda\approx433$. Although the dataset duration is insufficient to directly observe convergence of $h(t)\rightarrow h$, the assumption of stationarity enables statistical characterization of Lagrangian velocity gradient in the Protean frame~\cite{Lester:2025aa}, which permits synthetic generation of Lagrangian stretching variates and computation of $h(t)$, detailed as follows.

As the Lagrangian velocity gradient decorrelates according to the Kolmogorov time $\tau_v=\tau_\eta=\sqrt{\varepsilon/\nu}=0.0844$s~\cite{Li:2008aa}, sampling Lagrangian velocity gradient data for 10s over $10^3$ random tracer trajectories provides $\sim 10^5$ independent velocity gradient samples, sufficient for accurate statistical characterization. The $\epsilon_{11}$ component of the Protean velocity-gradient tensor follows a Laplace distribution with mean $\lambda_\infty=3.045$s$^{-1}$ and variance $\sigma_\epsilon=3.280$s$^{-1}$~\cite{Li:2008aa,Lester:2025aa}, yielding $\sigma_\infty=\sigma_\epsilon\sqrt{\tau_v}=0.953$s$^{-1}$. Using this Laplace distribution, we generate random $\epsilon_{11}$ variates to compute $N_0=10^3, 10^4$ stretching variates $\zeta_{k,n}=\zeta_i(n \tau_v)=\tau_v\sum_{k=1}^n\epsilon_{k,i}$ and compute the FTTE $h(t)=h(n \tau_v)=h_n=\ln[\frac{1}{N_0}\sum_{i=1}^{N_0}\exp(\zeta_{i,n})]/(n\tau_v)$ for $n=1:10^6$ time steps.

\begin{figure}
\centering
\includegraphics[width=0.3\columnwidth]{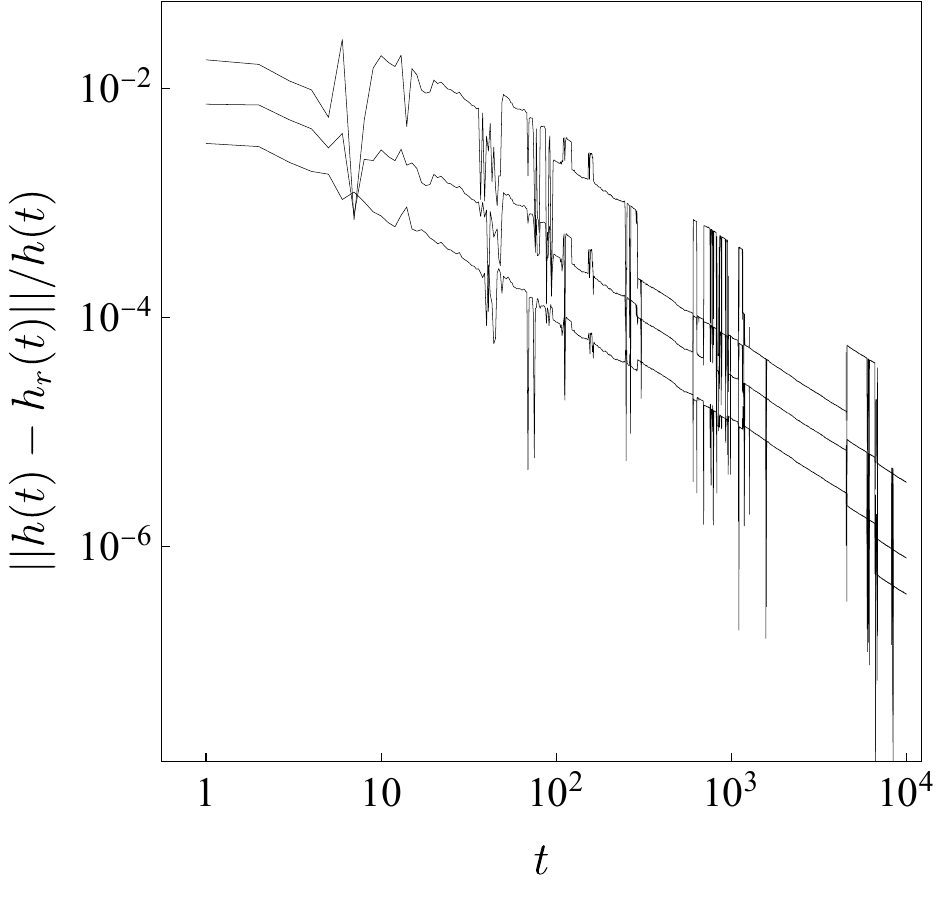}
\caption{Comparison of $h(t)$ computed from $N_0=10$ independent random stretching variates $\epsilon$ per time step (using the same statistics as used in Fig.~1b in the main paper) and replica entropy $h_r(t)$ computed from $\mathcal{N}=10^m$ ($m=2:4$, top to bottom) replicas generated from random combinations of the $N_0=10$ independent $\epsilon$ variates at each time step.}
\label{fig:replicas}
\end{figure}

In Fig.1b of the main paper we also compute an estimate for $h(t)$ for evolving $N(t)$ based on diffusive dispersion $\langle\Delta r(t)^2\rangle\sim t$, yielding $N(t)\sim V(t)=10^5+t^{3/2}\in[10^5,10^9]$, as the transition from Richardson to diffusive dispersion occurs for times greater than the large eddy turnover time $T_L\approx 1.99$ s~\cite{Li:2008aa}. Fig~\ref{fig:replicas} in the SM also uses this data to compare $h(t)$ computed from a small sample $N_0=10$ of independent $\epsilon$ variates at each time step with the \emph{replica} entropy $h_r(t)$ computed from $\mathcal{N}=10^m$, $m=2:4$ replicas generated from random combinations of the $N_0=10$ independent $\epsilon$ variates at each time step.

\subsection{Gaussian stretching}

Fig. 2a in the main paper also shows convergence of $h(t)$ based on statistical simulation of $h(t)$ with $N(t)=10^6+10t$ stretching variates (representing Fickian dispersion in a 2D flow), with $\tau_v=1$, and $\epsilon$ Gaussian distributed with mean $\lambda_\infty=1$, and standard deviation $\sigma_\epsilon=\sigma_\lambda=2^m$, $m=1:4$. The stretching variates $\zeta_{k,n}$ and $h(t)$ are computed as described above for homogeneous isotropic turbulence.

\subsection{Kraichnan flow}

We also consider a time-dependent Kraichnan flow which is generated according to
\begin{align}
    \mathbf u(\mathbf{X},t) = \sqrt{\frac{2}{N}} \sum\limits_{n=1}^N \cos(\mathbf k^{(n)} \cdot \mathbf{X} + \phi_n + \omega_n t) \mathbf p(\mathbf k^{(n)}),
\end{align}
where
\begin{align}
    p_1(\mathbf k) = \frac{k_2}{|\mathbf k|}, && p_2(\mathbf k) = \frac{-k_1}{|\mathbf k|}.
\end{align}
The $\mathbf k_j$ are sampled from a unit-Gaussian distribution. The phases $\phi_j$ are sampled from a uniform distribution between $0$ and $2 \pi$. The frequencies $\omega_j$ are sampled from a unit-Gaussian distribution. The flow field is divergence-free by construction. In the limit $N \to \infty$, $\mathbf u(\mathbf{X},t)$ is a multi-Gaussian random field with a Gaussian-shaped covariance function. Particle tracking is solved by an Euler-scheme,
\begin{align}
    \mathbf{X}(t + \Delta t) = \mathbf{X}(t) + \mathbf u[\mathbf{X}(t),t] \Delta t,
\end{align}
with $\Delta t = 10^{-2}$. The elongation of an infinitesimal strip is determined numerically as
\begin{align}
    \mathbf z(t + \Delta t) = \mathbf z(t) + \boldsymbol{\epsilon}(t) \cdot \mathbf z(t) \Delta t, 
\end{align}
where $\boldsymbol \epsilon(t)$ is given by
\begin{align}
    \epsilon_{ij}(t) &= (-1)^i \sqrt{\frac{2}{N}} \sum\limits_{n=1}^N \cos[\mathbf k^{(n)} \cdot \mathbf{X}(t) + \phi_n + \omega_n t] p_i(\mathbf k^{(n)}) k^{(n)}_{j}.
\end{align}
The FTTE $h(t)$ associated with stretching of line elements in this flow are reported in Fig.~2b of the main paper.

\section{Random walk stretching model}
In this section, we discuss the modeling of the stretching time series as a random walk process. First, we discuss the implications of Ito versus Stratonovic interpretation of the stochastic process characterized by a multiplicative noise. Then, we discuss the modeling of the line elongation using an Ornstein-Uhlenbeck process, which is compared to the data obtained from stretching in the Kraichnan flow discussed in the previous section. 

\subsection{Ito versus Stratonovic interpretation}

The elongation of an infinitesimal line element $\delta \ell(t,a)$ can be described by the stochastic differential equation
\begin{align}
\frac{d \delta \ell(t,a)}{dt} = \lambda_{\infty} \delta \ell(t,a) + \epsilon(t,a) \delta \ell(t,a). 
\end{align}
We model $\epsilon(t,a)$ as a Gaussian white noise with variance $\langle \epsilon(t,a) \epsilon(t',a) \rangle = \sigma_\infty^2 \delta(t - t')$. 
We consider now the process $\xi(t,a) = \ln \delta \ell(t,a)$. Using the Ito interpretation, its evolution is given by
\begin{align}
\frac{d \xi(t,a)}{dt} = \lambda_{\infty} - \frac{\sigma_\infty^2}{2} + \epsilon(t,a). 
\end{align}
When we use the Stratonovic interpretation, it is given by
\begin{align}
\frac{d \xi(t,a)}{dt} = \lambda_{\infty} + \epsilon(t,a),
\end{align}
Note that we use throughout the Ito calculus. In the first case, the distribution of $\xi(t)$ is
\begin{align}
p_\zeta(x,t) = \frac{\exp\left[- \frac{(x - \lambda_\infty t)^2}{2 \sigma_\infty^2 t} \right]}{\sqrt{2 \pi \sigma_\infty^2 t}}
\end{align}
and in the second case by
\begin{align}
p_\xi(x,t) = \frac{\exp\left[- \frac{\left(x - \lambda_\infty t + \frac{\sigma_\infty^2 t}{2} \right)^2}{2 \sigma_\infty^2 t} \right]}{\sqrt{2 \pi \sigma_\infty^2 t}}
\end{align}

Therefore, in the Ito interpretation, the topological entropy obtained from the ensemble average is $\lambda_\infty$, in the Stratonovic interpretation it is $\lambda_\infty + \sigma_\infty^2/2$. This can be seen by performing the ensemble average of $\delta \ell(t,a)$
\begin{align}
\langle \delta \ell(t,a) \rangle = \delta a \left\langle \exp\left[\xi(t,a)\right] \right\rangle. 
\end{align}
However, the physically meaningfull interpretation is the Stratonovic interpretation because it acknowledges the fact the white noise approximation for $\epsilon(t,a)$ is obtained as a limit process from a finite time correlation as explored in the next section.  

\subsection{Ornstein-Uhlenbeck model}

\begin{figure}
    \centering
    \includegraphics[width=0.45\linewidth]{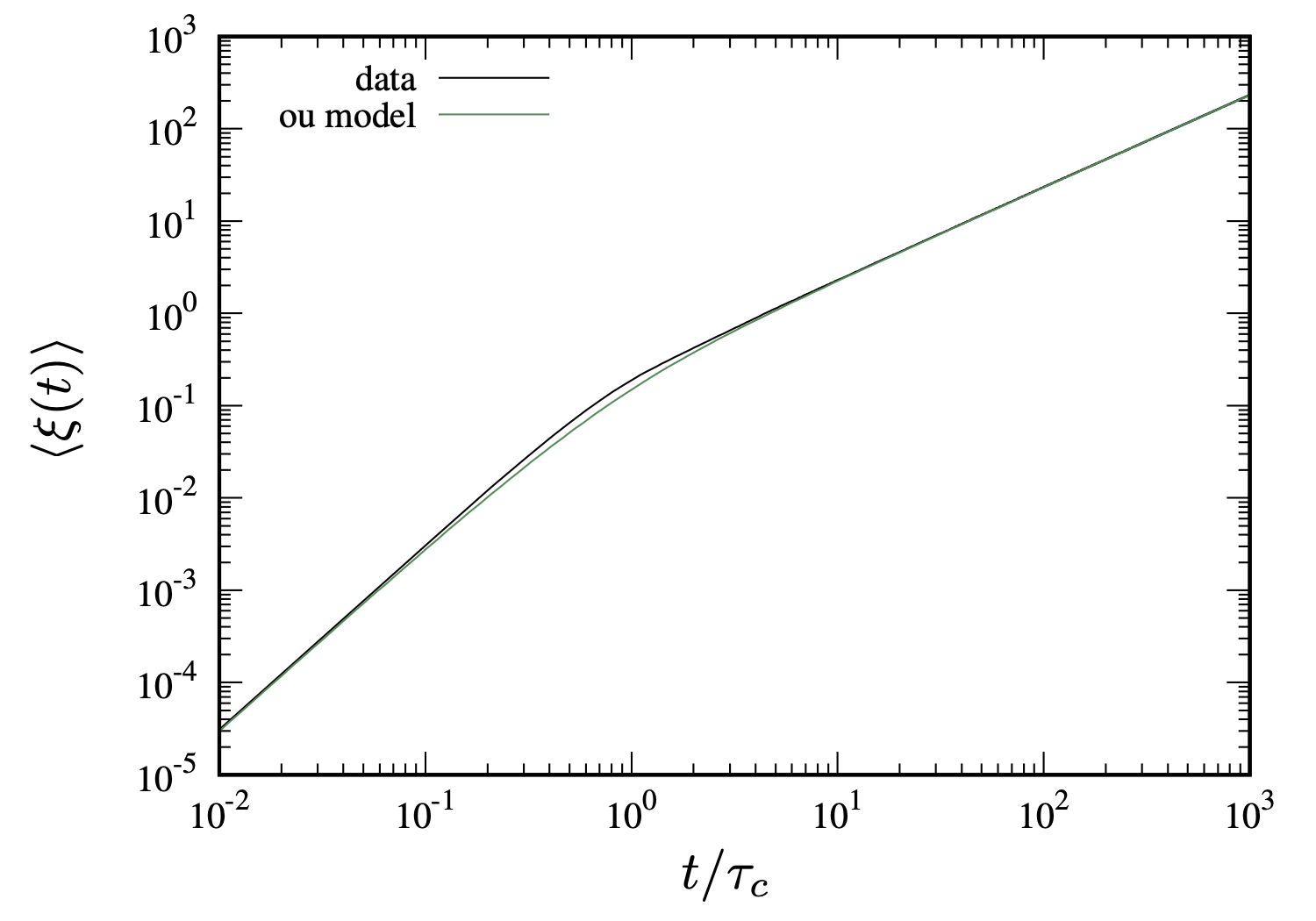}
    \includegraphics[width=0.45\linewidth]{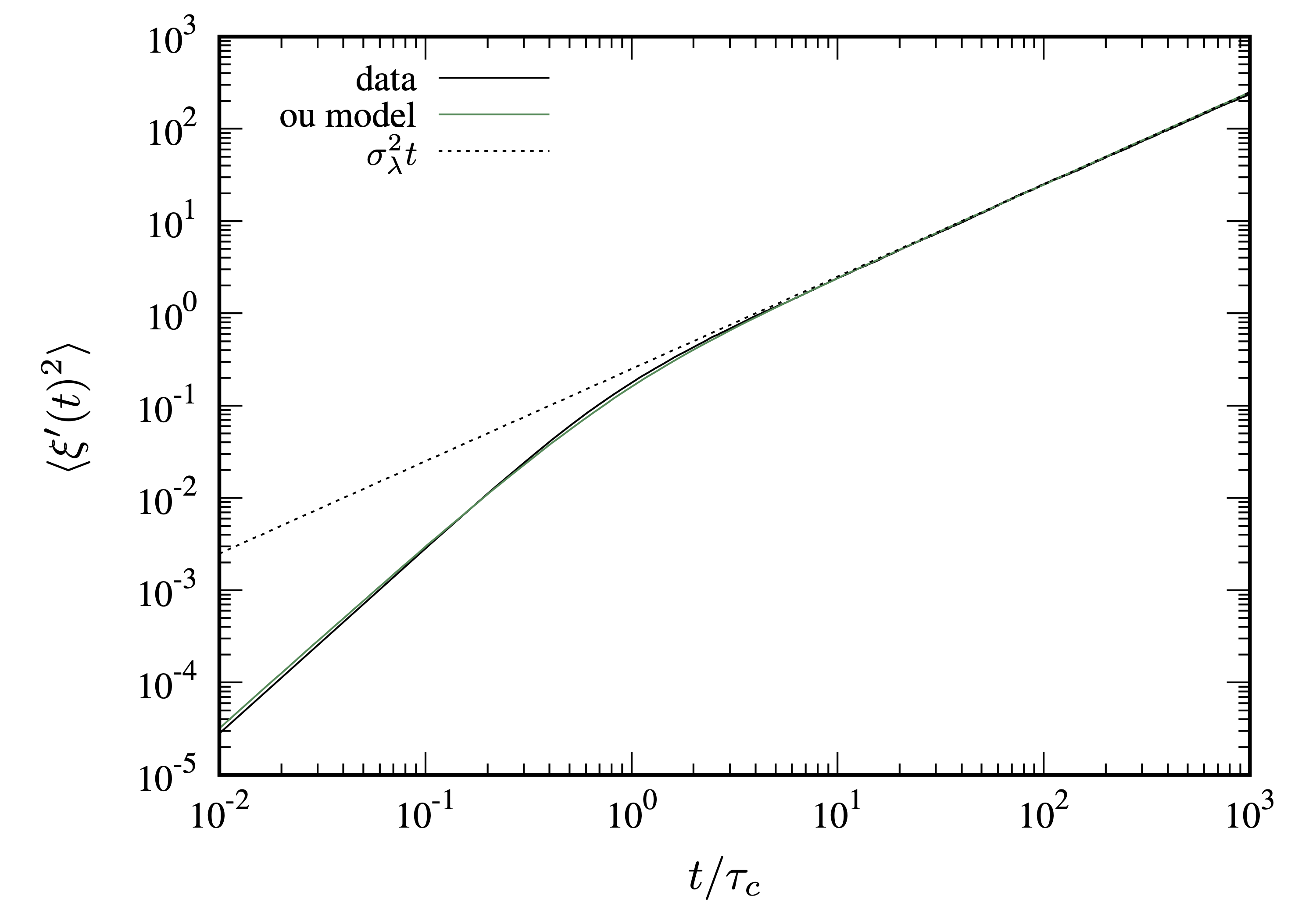}
    \caption{Comparison of mean $\langle\xi(t)\rangle$ and variance $\langle\xi'(t)^2\rangle$ of log-length $\xi(t)$ of an infinitesimal line element from numerical simulation of the Kraichnan flow and the analytic expressions (\ref{eqn:mean_xi}) (\ref{eqn:var_xi}) for the OU process (\ref{eqn:OU}).}
    \label{fig:OU}
\end{figure}

The evolution of the logarithm $\xi(t)$ of the line elongation is given by 
\begin{align}
\label{app:xi}
    \frac{d \xi(t)}{dt} = \epsilon(t). 
\end{align}
We model the deformation rate time series $\epsilon(t)$ as an Ornstein-Uhlenbeck (OU) process
\begin{align}
    \frac{d \epsilon(t)}{dt} = - \gamma [\epsilon(t) - \lambda_\infty] + \sqrt{2 \kappa} \zeta(t),\label{eqn:OU}
\end{align}
where $\gamma$ is the relaxation rate, $\zeta(t)$ a Gaussian white noise and $\kappa$ the fluctuation strength. The characteristic correlation scale is given by $\tau_v = \gamma^{-1}$. The mean deformation rate evolves as 
\begin{align}
    \langle \epsilon(t) \rangle = (\langle \epsilon_0 \rangle - \lambda_\infty) \exp(-\gamma t) + \lambda_\infty
\end{align}
We assume stationary initial conditions and set $\langle \epsilon_0 \rangle = 0$ and the initial variance equal to the stationary variance $\sigma_{\epsilon_0}^2 = \sigma_\epsilon^2$. Thus, the covariance function of the deformation fluctuation $\epsilon'(t) = \epsilon(t) - \lambda_{\infty}$ is given by
\begin{align}
    \langle \epsilon'(t) \epsilon'(t') \rangle = \sigma_\epsilon^2 \exp[-\gamma (t - t')], 
\end{align}
where $\sigma_\epsilon^2 = \kappa \gamma^{-1}$ and $t \geq t'$. The log-elongation $\xi(t)$ is obtained by integration of Eq.~\eqref{app:xi} as
\begin{align}
    \xi(t) = \int\limits_0^t dt' \epsilon(t). 
\end{align}
Thus, we obtain for the mean of $\xi(t)$
\begin{align}
    \langle \xi(t) \rangle = \lambda_\infty t - \frac{\lambda_\infty}{\gamma} \left[1 - \exp(- \gamma t) \right],\label{eqn:mean_xi}
\end{align}
and for the variance
\begin{align}
    \langle \xi'(t)^2 \rangle = \frac{2 \sigma_{\epsilon}^2}{\gamma^2} \left[\gamma t - 1 + \exp(-\gamma t) \right]. \label{eqn:var_xi}
\end{align}
The long-time evolution for $\gamma t \gg 1$ is given by 
\begin{align}
    \langle \xi'(t)^2 \rangle = \frac{2 \sigma_{\epsilon}^2}{\gamma} t \equiv \sigma_\infty^2 t.
\end{align}
Figure~\ref{fig:OU} shows that these results compare very well with numerical simulations of the 2D Kraichnan flow.
%

\section{Finite sample statistics}

The $\varepsilon'_{r,i}$ within a series $\{\varepsilon'_{r,i}\}_{i = 1}^n$ are mutually independent because for a given series the $\varepsilon'_{r,i} = \epsilon'_{k_i,i}$ with $k_i \in [1,N_0]$.  Expression \eqref{eq:ell_zeta1} indicates the the number of replica of $\zeta_{r,n}$ increases exponentially fast with $n$. However, the number of independent replica is not. In fact, the maximum number of independent replica of $\zeta_{r,n}$ is equal to $n N_0/n = N_0$. The nominator is due to the fact that after $n$ steps there is a total of $n N_0$ independent $\epsilon'_{k,i}$, from which the $\varepsilon'_{r,i}$ are sampled. The denominator comes from the fact that each $\zeta'_{r,n}$ is based on a series of $n$ different $\varepsilon'_{r,i}$, which need to be sampled without replacement from the set of $\epsilon'_{k,i}$ in order to have a set of independent $\zeta_{r,n}$. That is, out of the sample of $N_0^n$ replica, we can form $N_0^n/N_0 = N_0^{n-1}$ subsamples that have the same (finite) statistics. Thus, we can reorganize the sum and group the replica into $N_0^{n-1}$ families of independent replica such that 
\begin{align}
\label{eq:ell_zeta0b}
\ell_n = \ell_0 \exp(n\tau_v \lambda_\infty) \frac{1}{N_0^{n-1}} \sum\limits_{r=1}^{N_0^{n-1}} \frac{1}{N_0} \sum\limits_{k = 1}^{N_0} \exp\left({\zeta}^{(r)}_{k,n}\right)
\end{align}
Thus, we can write 
\begin{align}
\ell_n =  \ell_0 \exp(n \tau_v \lambda_\infty) \frac{1}{N_0^{n-1}} \sum\limits_{r=1}^{N_0^{n-1}} \int\limits_{-\infty}^\infty d x p_\zeta^{(r)}(x,n) \exp(x), 
\end{align}
where $p_\zeta^{(r)}(x,n)$ is the distribution of $N_0$ independent $\zeta^{(r)}_{k,n}$ within a family,
\begin{align}
p_\zeta^{(r)}(x,n) = \frac{1}{N_0} \sum\limits_{k = 1}^{N_0} \delta(x - \zeta^{(r}_{k,n}).   
\end{align}
Note that all families $\{\zeta^{(r)}_{k,n}\}_{k=1}^{N_0}$ share the same maximum because they are based on the same $\epsilon'$-statistics. Thus, as the (finite) statistical content of each family is the same, we set  $p_\zeta^{(r)}(x,n) = p_\zeta(x,n)$ independent of $r$.

Fig.~\ref{fig:replicas} in the SM shows that for $t \gg \tau_v$, $h(t)$ comprised of $N_0$ independent stretching variates $\epsilon$ at each time step is essentially identical to the \emph{replica} entropy $h_r(t)$ comprised of a large number $\mathcal{N}_n\gg N_0$ of replicas $\zeta_{r,n}$ composed from this set of independent $\epsilon$. 

\section{Path integral}
The expression for the elongation can be discretized on the time scale $\tau_v$ as 
\begin{equation}
        \label{eq:ellnb}
\ell_n = \ell_0 \int d \mathbf{X} \rho(\mathbf{X}) \exp(\tau_v \epsilon[\mathbf{x}(t_{n-1},\mathbf{X}),t_{n-1}]) \dots
\exp(\tau_v \epsilon[\mathbf{x}(t_0,\mathbf{X}),t_0]),
\end{equation}
where $t_n = n \tau_v$, $\ell_n = \ell(t_n)$. Note that $t_0 = 0$ and $\mathbf{X}(t = t_0,\mathbf{X}) = \mathbf{X}$. Eq. \eqref{eq:elln} can be written as 
\begin{align}
&\ell_n = \ell_0 \int d \mathbf{x}_{n-1} \dots \int d \mathbf{x}_0 \Big\{\int d \mathbf{X} \rho(\mathbf{X})\delta[\mathbf{x}_0 - \mathbf{x}(t_0,\mathbf{X})] \dots \delta[\mathbf{x}_{n-1} - \mathbf{x}(t_{n-1},\mathbf{X})]
\Big\}
\nonumber
\\
&\exp(\tau_v \epsilon(\mathbf{x}_0,t_0) \dots 
\exp(\tau_v \epsilon(\mathbf{x}_{n-1},t_{n-1})
\label{eq:elln}
\end{align}
The expression in the curly brackets is the joint distribution $g(\mathbf{x}_{n-1},t_{n-1};\dots;\mathbf{x}_0,t_0)$. 


\section{Derivation of Eq. (15) }

From equation (14) in the main paper

\begin{align}
\ell_n = \ell_0 \prod\limits_{i = 0}^{n-1} \int d\mathbf{x}_{i} \exp[\tau_v
\epsilon(\mathbf{x}_{i},t_{i})] g(\mathbf{x}_{i},t_{i}),
\end{align}
we assume that the line segments are uniformly distributed over $\Omega$, which implies
\begin{align}
\label{eq:elln_markov01}
\ell_n = \ell_0 \prod\limits_{i = 0}^{n-1} \frac{1}{V_0} \int d\mathbf{x}_{i} \exp[\tau_v
\epsilon(\mathbf{x}_{i},t_{i})].
\end{align}
Next, we subdivide the support volume $\Omega$ into $k=1:N_0$ stretching regions of volume $\Delta V_k$ (where $\Delta V_k=\ell_k^d$ for mono-scale flows)
\begin{align}
\label{eq:elln_markov01}
\ell_n = \ell_0 \prod\limits_{i = 0}^{n-1} \frac{1}{V_0} \sum_{k=1}^{N_0} \Delta V_k \exp[\tau_v
\epsilon_{k,i}].
\end{align}
and assume that the $\Delta V_k$ are all of approximately the same size $\Delta V_k$ (which can be relaxed for multi-scale flows and would imply that the different regions are weighted differently, however this does not impact the main results) such that we obtain Eq. (15).

\section{Derivation of Eq. (25)}

%
%
%
%
We decompose the initial volume $\Omega(t_0)$ into $N_0$ stretching regions $\Omega_k(t_0)$ and assume that the line segments that are in the same initial subdomain $\Omega_k(t_0)$ do not disperse into other stretching regions. Thus, we write 
\begin{align}
    g(\vx_{n-1},t_{n - 1},\dots,\vx_0, t_0) = \sum\limits_{k = 1}^{N_0} \mathbb I[\vx_{n-1} \in \Omega_k(t_{n-1})] \dots \mathbb I[\vx_{0} \in \Omega_k(t_0)] \frac{\Delta V_k}{V_0},
\end{align}
that is, the positions of the line segments are propagated within the same subdomain $\Omega_k(t)$ of constant volume $\Delta V_k$. Inserting this expression into Eq. (13) in the main text gives
\begin{align}
\label{sm:eq:elln_markov-1}
        \ell_n = \ell_0 & \sum\limits_{k = 1}^{N_0} \frac{\Delta V_k}{V_0} \exp(\tau_v \epsilon_{k,n-1}) \dots
\exp(\tau_v \epsilon_{k,0}),
\end{align}
where we used that the $\epsilon(\vx_i,t_i) \equiv \epsilon_{k,i}$ for $\vx_i \in \Omega_k(t_i)$ are uniform within a stretching region. As above, we assume for simplicity that $\Delta V_k = \Delta V_0$ and so 
\begin{align}
\label{sm:eq:elln_markov-2}
        \ell_n = \ell_0 & \frac{1}{N_0}\sum\limits_{k = 1}^{N_0} \exp(\tau_v \epsilon_{k,n-1}) \dots
\exp(\tau_v \epsilon_{k,0}).
\end{align}
Inserting $\epsilon_{k,i} = \lambda_\infty + \epsilon'_{k,i}$ and using the definition of $\zeta_{k,n} = \sum\limits_{i=1}^n \tau_v \epsilon'_{k,i}$ gives Eq. (25) in the main text. 

\bibliography{reflist}